\documentclass[aps,prl,twocolumn,showpacs,letterpaper]{revtex4-1}
\pdfoutput=1
\usepackage{epsfig}
\usepackage{amsmath}
\usepackage{amssymb}
\usepackage{wasysym}
\usepackage[colorlinks=true,linkcolor=blue, citecolor=red]{hyperref}

\newcommand{\bs}{\boldsymbol}

\begin{document}

\title{Bound States of Conical Singularities in Graphene-Based Topological Insulators}
\author{Andreas R\"uegg}
\affiliation{Department of Physics, University of California, Berkeley, California 94720, USA}
\author{Chungwei Lin}
\affiliation{Department of Physics, The University of Texas at Austin, Austin, Texas 78712, USA}

\date{\today}

\begin{abstract}
We investigate the electronic structure induced by wedge-disclinations (conical singularities) in a honeycomb lattice model realizing Chern numbers $\gamma=\pm 1$. We establish a correspondence between the bound state of (i) an isolated $\Phi_0/2$-flux, (ii) an isolated pentagon $(n=1)$ or heptagon $(n=-1)$ defect with an external flux of magnitude $n\gamma \Phi_0/4$ through the center and (iii) an isolated square or octagon defect without external flux, where $\Phi_0=h/e$ is the flux quantum. Due to the above correspondence, the existence of isolated electronic states bound to disclinations is robust against various perturbations. Hence, measuring these defect states offers an interesting probe of graphene-based topological insulators which is complementary to measurements of the quantized edge currents.
\end{abstract}



\maketitle

The surface states of topological insulators (TIs) \cite{Moore:2010,Hasan:2010,Qi:2011b} are protected by time-reversal symmetry and charge conservation, both of which can persist independently from microscopic details. However, in sufficiently pure materials, crystalline symmetries can equally protect non-trivial properties of the electronic structure \cite{Mong:2010,Hughes:2011,Fu:2011}, such as surface states of certain high-symmetry surfaces \cite{Dziawa:2012,Xu:2012}. That the presence of crystalline symmetries enriches the topological response is further exemplified by the observation that dislocations in the crystal lattice can robustly bind in-gap states in certain TIs \cite{Ran:2009,Teo:2010,Ran:2010,Juricic:2012}. In these instances, the electrons near the Fermi energy acquire a Berry phase of $\pi$ when encircling the defect which induces changes of the electronic structure in analogy to a magnetic flux tube with half-integer multiple of the flux quantum $\Phi_0=h/e=2\pi$. Namely, in 2D, a single Kramer's pair appears in the gap \cite{Lee:2007,Ran:2008,Qi:2008,Ruegg:2012,Assad:2012} while in 3D, protected one-dimensional modes form \cite{Zhang:2009,Rosenberg:2010b,Peng:2010,Bardarson:2010,ZhangYi:2010}.

\begin{figure}[h!]
\includegraphics[width=0.9\linewidth]{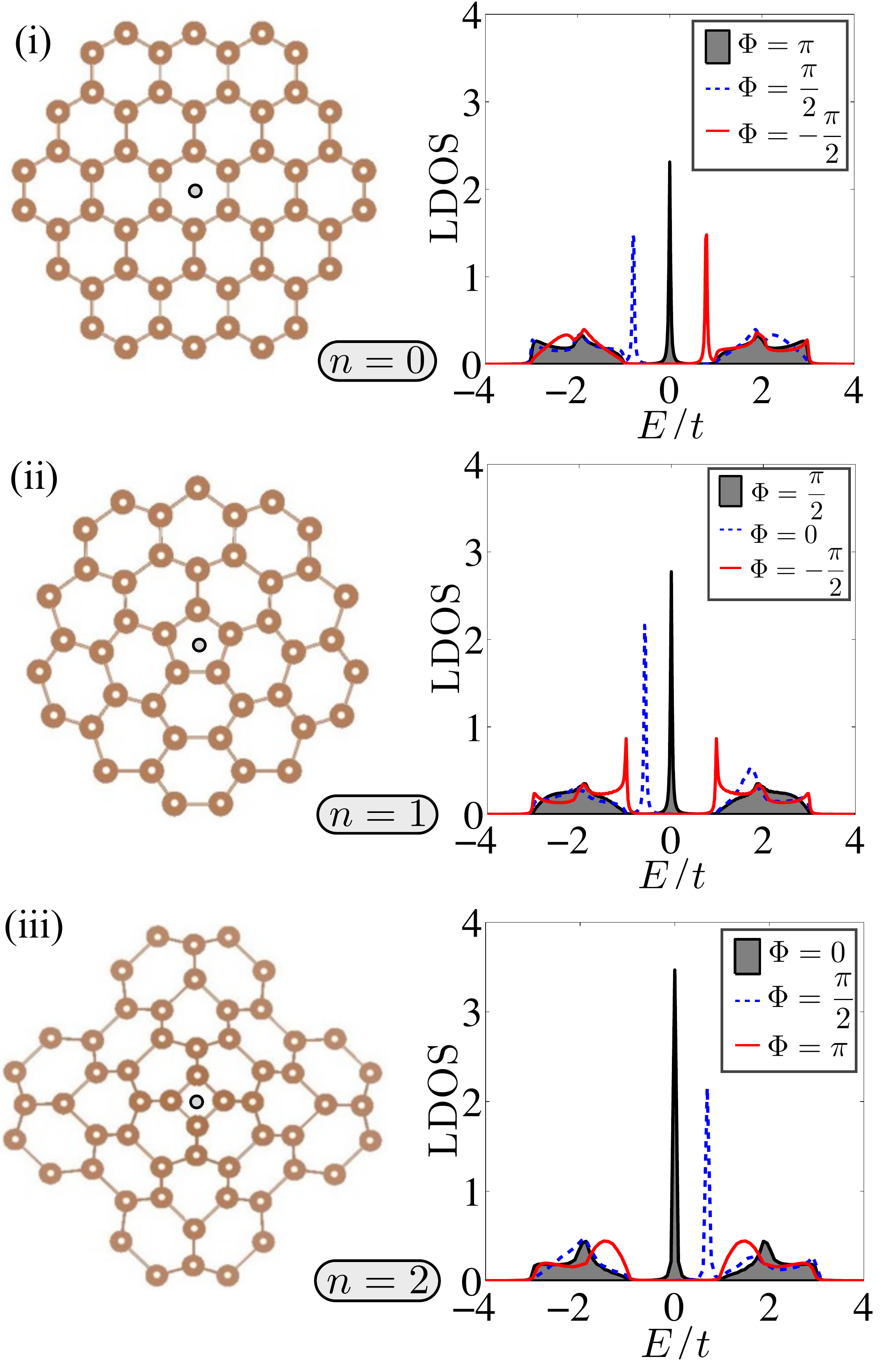}
\caption{Correspondence between the bound state of (i) and isolated $\pi$ flux in the defect-free case $(n=0)$, (ii) an isolated pentagon defect $(n=1)$ with an external flux $\pi/2$ and (iii) an isolated square defect $(n=2$) without external flux. The external flux $\Phi$ was applied through the center marked by a circle. Results are presented for the model Eq.~\eqref{eq:Haldane} with $t_{2i}/t=0.4$. The local density-of-states (LDOS) were obtained on a site of the central polygon using the Lanczos algorithm.}
\label{fig:disclinations}
\end{figure}

In this Letter, we demonstrate that also disclinations can robustly bind in-gap states in certain TIs with crystalline symmetries. Our conclusion is based on the study of wedge disclinations on the honeycomb lattice, see Fig.~\ref{fig:disclinations}. Such conical defects form the elementary building blocks of various extended lattice defects observed in graphene and related carbon based structures \cite{Gonzalez:1993,Lammert:2000,Lammert:2004,Cortijo:2007,Yazyev:2010,Cockayne:2011}. While the intrinsic spin-orbit coupling in graphene is too small to access the TI phase \cite{Kane:2005a, Kane:2005b} experimentally, several promising routes exist to stabilize a topological phase, either by enhancing the intrinsic spin-orbit coupling via adsorption \cite{CastroNeto:2009,Weeks:2011, Ding:2011,Hu:2012} or by using Rashba spin-orbit coupling \cite{Tse:2011,Zhenhua:2011}. Our discovery of robust defect states demonstrates the possibility of a probe of the topological state in graphene-based TIs or related systems \cite{Gomes:2012,LiuC:2011,Vogt:2012}, which is complementary to the measurement of quantized edge currents.
 
On the hexagonal lattice, an isolated wedge disclination is constructed by locally replacing a hexagon by a $f$-gon (we discuss $f=4,5,7,8$) while preserving the three-fold connectivity of the honeycomb lattice. This introduces a global change of the lattice best illustrated by Volterra's cut-and-glue construction \cite{Chaikin:2000}, in which a wedge is removed from or added before gluing the two sides back together to form a cone. The point group symmetry restricts the possible opening angles to multiples of $\pi/3$ and we label different defects by the integer $n$ counting the number of removed ($n>0)$ or added ($n<0$) $\pi/3$ wedges. To study the interplay between such conical singularities and electrons in topologically non-trivial bands, we investigated a model of a Chern insulator for spinless fermions on the honeycomb lattice, first introduced by Haldane \cite{Haldane:1988}:
\begin{equation}
{\mathcal H}=-t\sum_{\langle i,j\rangle}\left(c_{i}^{\dag}c_{j}+{\rm h.c}\right)+t_{2i}\sum_{\langle\langle i,j\rangle\rangle}\left(i\nu_{ij} c^{\dag}_{i}c_{j}+{\rm h.c}\right).
\label{eq:Haldane}
\end{equation}
The real nearest-neighbor hopping is $t$ and we assume a purely imaginary second-neighbor hopping $it_{2i}\nu_{ij}$ where $\nu_{ij}=\pm1$ depends on the hopping direction \cite{Haldane:1988}. At half-filling, the model defined in Eq.~\eqref{eq:Haldane} has a finite Hall conductivity $\sigma_{xy}=\gamma e^2/h$ with a Chern number $\gamma={\rm sign}(t_{2i})$. A generalization to a time-reversal invariant topological insulator, in which the imaginary second neighbor hopping is generated by intrinsic spin-orbit coupling, has been discussed by Kane and Mele \cite{Kane:2005a,Kane:2005b} and our results generalize to this situation, as well.

The main findings of this work is the connection between the bound state induced by (i) an isolated $\Phi_0/2$-flux, (ii) an isolated pentagon $(n=1)$ or heptagon $(n=-1)$ defect with an external flux of magnitude $n\gamma \Phi_0/4$ through the center and (iii) an isolated square $(n=2)$ or octagon $(n=-2)$ defect without external flux. We reached this conclusion in three different ways: (I) direct computation of the local density of states in the lattice model, (II) the analysis of disclinations in the continuum model and (III) their description in terms of coupled edge modes.

To compute the local density of states (LDOS) near the defect core in the lattice model, we fixed the ratio $t_{2i}/t=0.4$ and used the Lanczos algorithm with open boundary condition \cite{Grosso:1985} to obtain the retarded local Green's function $G_{ii}(E)$ from which the LDOS $N_{i}(E)=-\frac{1}{\pi}{\rm Im}G_{ii}(E)$ was derived. We keep up to 300 states, and use a small imaginary part of 0.02 $t$ to obtain a smooth spectra. Figure~\ref{fig:disclinations}(i) shows the LDOS in the defect-free case on the hexagon through which an external flux is threaded. Turning on a finite flux \cite{note:supplementary} moves a bound state from the valence to the conduction band reaching $E=0$ for $\Phi=\Phi_0/2=\pi$. From particle-hole symmetry, $N_i(E)=N_i(-E)$, and the conservation of states, $\int_{-\infty}^{\infty}dE N_{i}(E)=1$, it follows that the excess or deficit charge bound to the $\pi$ flux is $\pm e/2$ \cite{Lee:2007}. Our numerical integration of the LDOS confirmed this expectation.

Studying the LDOS on a pentagon defect, Fig.~\ref{fig:disclinations}(ii), we identify an in-gap state even without external flux. Threading $\Phi=\pi/2$ through the pentagon shifts the bound state energy producing a mid-gap state in analogy to the situation (i) with $\pi$ flux. On the other hand, a flux $\Phi=-\pi/2$ produces two symmetric resonances close to the band edges. Switching from the pentagon to the heptagon defect $(n=-1)$ or changing the sign of the Chern number, we find that the opposite sign of the flux is required to produce the mid-gap state.

Figure~\ref{fig:disclinations}(iii) illustrates the case of a square defect. The LDOS shows that the mid-gap state is now realized in the absence of any external flux. To completely remove the bound state, an external $\pi$-flux is required. We find the same behavior also for the octagon defect (not shown). Moreover, this property does not rely on particle-hole symmetry (or the fact that the bound-state energy is at $E=0$): in an analogous calculation including also real second-neighbor hopping, we find that only an external $\pi$-flux is able to completely remove the bound state. The robustness of the correspondence between (i), (ii) and (iii) is further discussed below.

The numerical results presented in Fig.~\ref{fig:disclinations} can be consistently explained from a continuum description, as we discuss in the following. In the low-energy limit, Eq.~\eqref{eq:Haldane} reduces to the Dirac Hamiltonian with a ``Haldane mass" $m=3\sqrt{3} t_{2i}$:
\begin{equation}
H=v[\tau_z\sigma_xp_x+\sigma_yp_y]+m\tau_z\sigma_z,
\label{eq:Dirac}
\end{equation}
where $\vec{\sigma}=(\sigma_x,\sigma_y,\sigma_z)$, $\vec{\tau}=(\tau_x,\tau_y,\tau_z)$ are Pauli matrices denoting the sublattice and valley degrees of freedom, respectively, and $v=\sqrt{3}ta/(2\hbar)$ is the Fermi velocity. $H$ acts on the four-component spinor $\Psi({\bs r})=[\psi_{A}({\bs r}),\psi_{B}({\bs r}),\psi_{A'}({\bs r}),\psi_{B'}({\bs r})]^T$ where $A$ and $B$ label the sublattice in valley $K$ and $A'$ and $B'$ in valley $K'$ \cite{note:supplementary}. It is known that deformations of the honeycomb lattice enter the continuum description via fictitious gauge fields \cite{Vozmediano:2010,Guinea:2010,Ghaemi:2012}. As we review below, topological point defects manifest themselves by spatially well-localized fluxes of the fictitious fields \cite{Gonzalez:1993}.

\begin{figure}[http]
\includegraphics[width=0.8\linewidth]{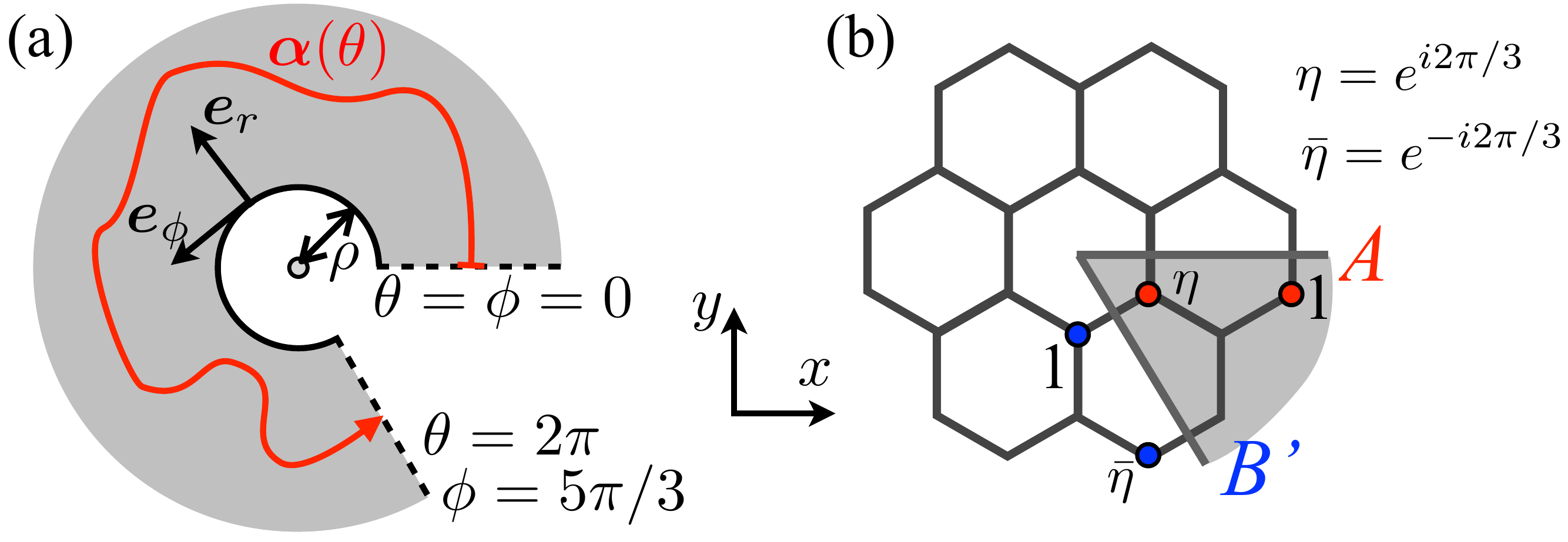}
\caption{(a) Continuum version of the cut-and-glue construction with a regularization hole of radius $\rho$ around the origin. ${\bs\alpha}(\theta)$ is a closed path around the cone. (b) The boundary conditions for the spinor across the seam have to compensate the mismatch of the base functions, as indicated for a pair of matching degrees of freedom $A$ and $B'$ for the pentagon disclination.}
\label{fig:continuum}
\end{figure}

We model the disclination by a regularized cone where a disk of radius $\rho$ around the apex is removed, see Fig.~\ref{fig:continuum}(a). The fictitious gauge fields are related to the non-trivial holonomy when the spinor is parallel transported along a closed path ${\bs \alpha}(\theta)$ ($0\leq\theta\leq 2\pi$) around the cone:
\begin{equation}
\Psi({\bs \alpha}(2\pi))=\mathcal{H}_n\Psi({\bs\alpha}(0)),\quad \mathcal{H}_n=e^{i\frac{\pi n}{3}\left(\frac{\sigma_z\tau_z-3\sigma_y\tau_y}{2}\right)}.
\label{eq:UnBC}
\end{equation}
$\mathcal{H}_n$ simply represents the rotation by the Frank angle $n\pi/3$ of the defect. The boundary condition for the envelope function, Eq.~\eqref{eq:UnBC}, compensates the mismatch of the base functions $e^{\pm i{\bs K}\cdot{\bs r}}$ across the seam in the cut-and-glue procedure illustrated in Fig.~\ref{fig:continuum}(b), making the total wave function single-valued \cite{Gonzalez:1993, Lammert:2000,Vozmediano:2010,Mesaros:2010,note:supplementary}. 

To deal with Eq.~\eqref{eq:UnBC} for general $n$, we seek for a local gauge in which the boundary condition is {\it independent} of disclination type. To achieve this goal, we introduce polar coordinates $(r,\phi)$ defined in the unfolded plane, see Fig.~\ref{fig:continuum}(a), and perform two singular gauge transformations $\Psi\overset{U}{\mapsto}\tilde{\Psi}\overset{V_n}{\mapsto}\tilde{\Psi}_n$ with
\begin{equation}
U(\phi)=e^{i\frac{\phi}{2}\sigma_z\tau_z}, \quad V_n(\theta)=e^{i \frac{n\theta}{4}\sigma_y\tau_y}.
\label{eq:Psi_n}
\end{equation}
The first operation $U$ transforms $\Psi$ to a co-rotating spinor \cite{note:co-rotating}, effectively replacing $\partial_r$ by $\partial_r+1/(2r)$ in the Hamiltonian. The second gauge transform $V_n$ introduces a matrix-valued gauge field into the Hamiltonian, effectively replacing $\partial_{\theta}$ by $\partial_{\theta}-i\frac{n}{4}\sigma_y\tau_y$. The transformed spinor $\tilde{\Psi}_n(r,\theta)$, $\theta=\phi/(1-\frac{n}{6})$, is now anti-periodic in $\theta$ for $any$ $n$.
In the final step, we use a {\it global} transformation $S$
\begin{equation}
\Psi_n'(r,\theta)=S\tilde{\Psi}_n(r,\theta),\quad S=\frac{1}{\sqrt{2}}(1+i\tau_x\sigma_y),
\end{equation}
which block-diagonalizes the Hamiltonian and defines two emergent valleys $\tau=\pm$. The separation into two decoupled valleys is well-known from the massless case \cite{Gonzalez:1993, Lammert:2000, Lammert:2004} but is {\it not} always possible in the presence of a mass term  \cite{note:valley_comments}. However, it is possible here for all types of disclinations because the Haldane mass $m\tau_z\sigma_z$ preserves the six-fold rotation symmetry around the center of a hexagon. The block-diagonal Hamiltonian $H_n'={\mathcal U}H{\mathcal U}^{\dag}$  
with ${\mathcal U}=SV_nU$, has the same form for any $n$.
The product {\it ansatz}
$
\Psi'_n(r,\theta)=\chi(r)e^{ij\theta}
$
with half-integer $j$
decouples radial and angular part and the radial part is $(\hbar=1=v)$ \cite{note:supplementary}
\begin{equation}
H'_{\tau}(n)=\frac{-i}{r}\left[\left(r\partial_r+\frac{1}{2}\right)\tau\sigma_x+i\nu_{\tau}(n)\sigma_y\right]+m\tau\sigma_z.
\label{eq:Hr}
\end{equation}
As before, $\tau=\pm$ denotes the emergent valley and \cite{Lammert:2000,Lammert:2004}
\begin{equation}
\nu_{\tau}(n)=\frac{j+\frac{\Phi}{\Phi_0}+\frac{n}{4}\tau}{1-\frac{n}{6}}.
\label{eq:nu}
\end{equation}
Equation~\eqref{eq:nu} also accounts for a localized real magnetic flux $\Phi$ through the origin \cite{note:supplementary,Slobodeniuk:2010}. The topological defect manifests itself through the denominator $1-n/6$ and an additional gauge flux of $n\tau\Phi_0/4$ with opposite sign in the two emergent valleys.

\begin{figure}[http]
\includegraphics[width=1\linewidth]{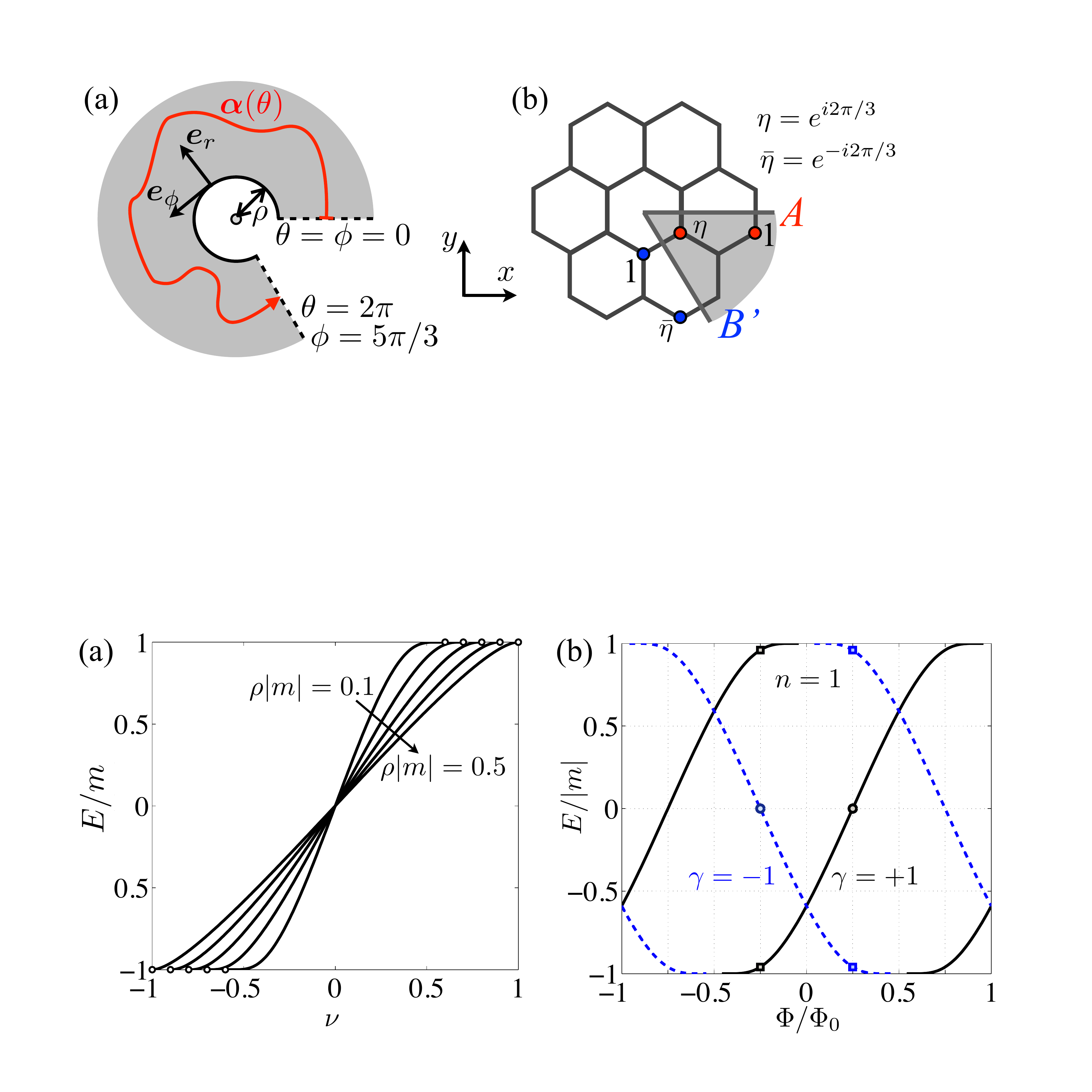}
\caption{(a) General solution for the bound-state energy from Eq.~\eqref{eq:quantization} as function of $\nu$ for different radii of the hole $\rho|m|$. (b) Bound-state energies as function of external flux $\Phi/\Phi_0$ for the pentagon defect.}
\label{fig:bound_state_energy}
\end{figure}

The eigenvalue problem $H'_{\tau}(n)\chi_{\tau}(r)=E\chi_{\tau}(r)$ can be solved in each valley separately. Bound states are given in terms of modified Bessel functions of the second kind which decay exponentially for $r\rightarrow\infty$. We find\begin{eqnarray}
\chi_{+}(r)=\begin{pmatrix}
K_{(\nu_+-1/2)}(\kappa r)\\
i\frac{\kappa}{m+E}K_{(\nu_++1/2)}(\kappa r)
\end{pmatrix},\\
\chi_{-}(r)=\begin{pmatrix}
-i\frac{m-E}{\kappa}K_{(\nu_-+1/2)}(\kappa r)\\
K_{(\nu_--1/2)}(\kappa r)
\end{pmatrix},
\label{eq:chi}
\end{eqnarray}
where $\kappa=\sqrt{m^2-E^2}>0$. 
The square integrability of $\Psi$ for $\rho\rightarrow 0$ does {\it not} uniquely determine the bound state \cite{Roy:2010}. To obtain quantized solutions, 
the internal structure of the disk $r<\rho$ has to be specified. The correct quantization is achieved by replacing the Haldane mass in Eq.~\eqref{eq:Hr} by a confining potential $V(r<\rho)=-M\sigma_z$ \cite{Berry:1987,Recher:2007,Akhmerov:2008}. The mass term $-M\sigma_z$, as compared to the Haldane mass, has opposite sign in one of the emergent valleys, thereby defining the topologically trivial insulator.
Because ${\mathcal U}^{\dag}\sigma_z{\mathcal U}=\sigma_x\tau_x$, we identify $V(r<\rho)$ in the frame of Eq.~\eqref{eq:Dirac} with the inversion symmetric mass term of a kekule distortion \cite{Mudry:2007,note:kekule_mass}. For $M\rightarrow+\infty$, matching of the wave function at $r=\rho$ takes the form
\begin{equation}
\gamma\left({\bs e}_{\phi}\cdot\hat{{\bs I}}\right)\Psi(\rho,\phi)=\Psi(\rho,\phi).
\label{eq:chiralBC}
\end{equation}
Here, $\hat{\bs I}=(-\tau_x\sigma_z,-\tau_y)^T$ is the normalized axial current in the frame of Eq.~\eqref{eq:Dirac} and ${\bs e}_{\phi}$ the azimuthal unit vector. The sign on the left-hand-side is fixed by the Chern number $\gamma={\rm sign}(m)$. For $\rho\rightarrow 0$, Eq.~\eqref{eq:chiralBC} realizes a special case of the general four parameter family of self-adjoint boundary conditions \cite{Roy:2010}. For the spinors of the form Eq.~\eqref{eq:chi}, it can only be satisfied in one valley. Moreover, Eq.~\eqref{eq:chiralBC} leads to the quantization of the bound-state energy through
\begin{equation}
\sqrt{\frac{m-E}{m+E}}=\frac{K_{\nu_{\gamma}-1/2}(\kappa\rho)}{K_{\nu_{\gamma}+1/2}(\kappa\rho)},
\label{eq:quantization}
\end{equation}
which, in combination with Eq.~\eqref{eq:nu}, incorporates the main result of the present work.
As illustrated in Fig.~\ref{fig:bound_state_energy}(a) for different values of the dimensionless radius $\rho|m|$, Eq.~\eqref{eq:quantization} has monotonic real solutions for the bound-state energy $|E|<|m|$ as function of $\nu$ in a range $|\nu|<1/2+\rho|m|$. For $\nu=0$ it follows that $E=0$ independent of $\rho |m|$. The physical bound-state spectrum as function of flux $\Phi/\Phi_0$ for a specific defect is constructed from the general solution by use of Eq.~\eqref{eq:nu} and is found to agree with the numerical results obtained in the lattice model. The case of a pentagon defect ($n=1$) with either sign of the Chern number $\gamma=\pm1$ is illustrate in Fig.~\ref{fig:bound_state_energy}(b). In particular, this solution predicts that insertion of an external magnetic flux $\gamma \pi/2$ [marked with $\circ$] shifts the bound state to zero energy while the opposite flux $-\gamma \pi/2$ [marked with ${\scriptscriptstyle \square}$] leads to two bound states symmetrically arranged with respect to $E=0$, in accordance with the results shown in Fig.~\ref{fig:disclinations}(ii).
%

\begin{figure}
\includegraphics[width=1.0\linewidth]{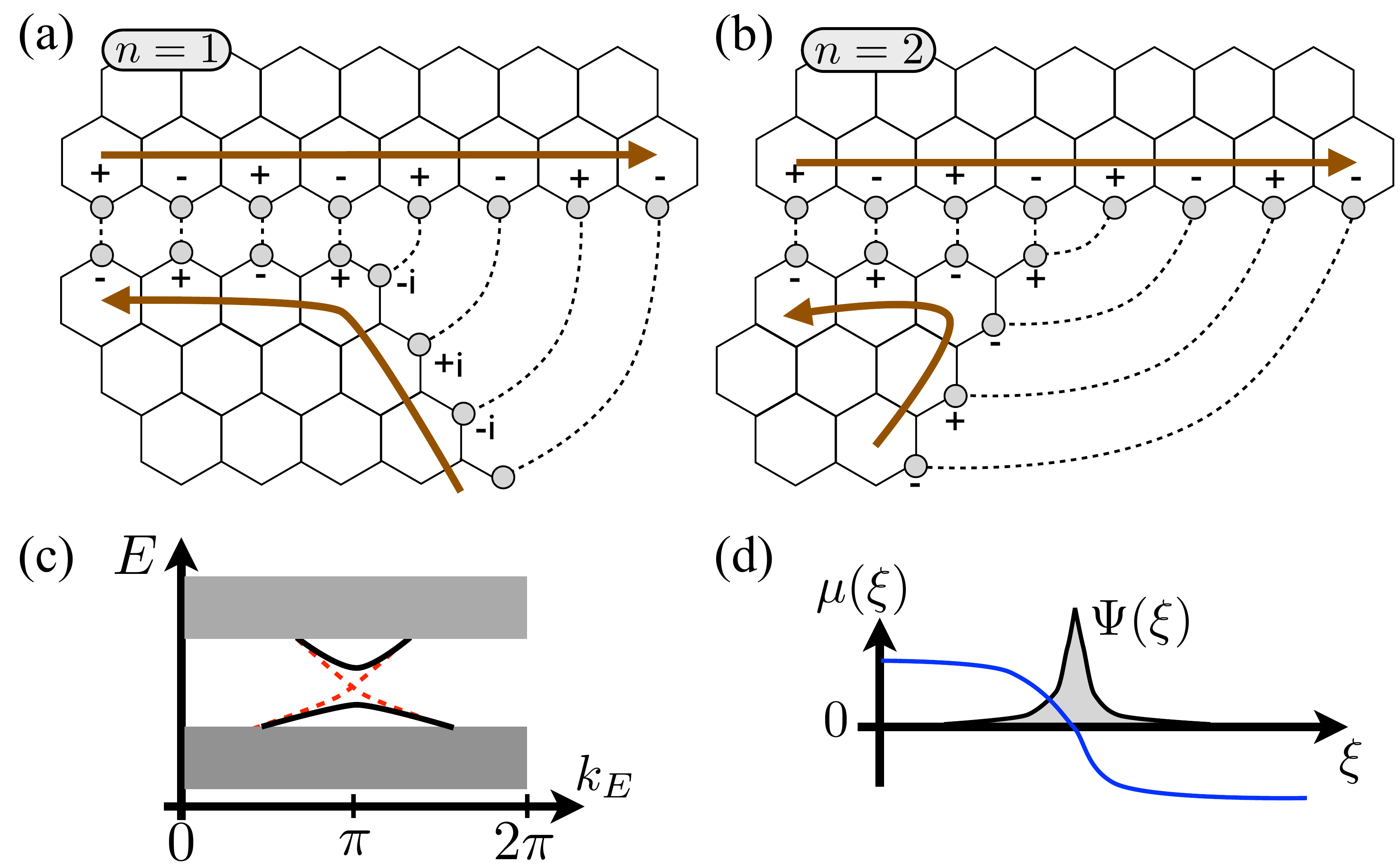}
\caption{Coupling of chiral edge states across the seam in the cut and glue construction of (a) a $n=1$ and (b) a $n=2$ disclination. (c) Without coupling, the edge states cross at $k_Ea=\pi$, which is protected by inversion symmetry. Turning on a weak coupling locally opens a gap. The different matching conditions between the edge states on the left- and right-hand side of the defect is described by a $\xi$-dependent mass term $\mu(\xi)$ where $\xi$ is the coordinate along the edge. (d) For $n=2$, $\mu(\xi)$ changes sign, implying a bound state.}
\label{fig:edges}
\end{figure}

The correspondence between external magnetic and internal fictitious fluxes induced by wedge disclinations has an intuitive explanation via the coupling of edge modes across the seam \cite{Ran:2009,Ran:2010}, as shown in Fig.~\ref{fig:edges}(a) and (b) for $n=1$ and $2$, respectively \cite{note:internal}.  We start with two disconnected flat honeycomb sheets from which a $60^{\circ}$- or $120^{\circ}$-wedge has been removed. The bulk-edge correspondence for Chern insulators implies chiral edge states propagating along the zig-zag edges of top and bottom part. In the vicinity of the energy crossing, they are described by the edge theory
\begin{equation}
H_{\rm edge}=\int\!d\xi\varphi^{\dag}(\xi)[-iv\partial_{\xi}\sigma_z+\mu(\xi)\sigma_x]\varphi(\xi)
\label{eq:Hedges}
\end{equation}
with $\mu(\xi)=0$. The two-component wave function $\varphi(\xi)$ varies smoothly on the scale of the lattice constant and includes right and left movers $\varphi=(\varphi_R,\varphi_L)^T$; $\xi$ is the coordinate along the cut. The total edge wave function on a lattice site is given by $\varphi_{\rm edge}(\xi)=e^{ik_E\xi}\varphi_R(\xi)+e^{-ik_E\xi}\varphi_L(\xi)$ where $k_E$ is the edge momentum. Inversion symmetry implies that the edge states of a zig-zag edge cross at either $k_Ea=0$ or $\pi$ - in the model Eq.~\eqref{eq:Haldane}, they cross at $k_Ea=\pi$, see Fig.~\ref{fig:edges}(c). Hence, the base functions $e^{\pm ik_E\xi}$ oscillates with a period of two and their amplitudes are indicated in Fig.~\ref{fig:edges}(a) and (b). A weak coupling between left and right movers is described by $\mu\neq0$ in Eq.~\eqref{eq:Hedges}. This gluing across the seam locally opens a gap of order $\mu$ in the edge spectrum. However, to account for the different matching conditions of the base functions, $\mu(\xi)$ acquires an additional factor $(\gamma i)^{n}$ on the right-hand side of the defect. For $n=2$, this corresponds to a sign change of $\mu(\xi)$ implying a bound state, cf.~Fig.~\ref{fig:edges}(d), in analogy to solitons in polyacetylene \cite{Su:1979}. This sign change is equivalent to a $\pi$ flux in the defect-free system \cite{Lee:2007}. Similarly, the factor $(\gamma i)$ for $n=1$ relates to a flux $\gamma\pi/2$.

If inversion symmetry is broken, the edge states cross away from $k_E a=\pi$. However, as long as the edge state theory can be obtained by expansion with the base functions at $k_Ea=\pi$, the correspondence between Frank angle and fictitious flux remains, even though the bound-state energy in general shifts. We have numerically confirmed this expected robustness by adding both local perturbations in the form of on-site potentials as well as various global symmetry-breaking terms including staggered sublattice potentials, real second neighbor hopping as well as dimerized first-neighbor hopping. 

Our main results are summarized in Fig.~\ref{fig:disclinations} and given by Eq.~\eqref{eq:quantization} in combination with Eq.~\eqref{eq:nu} which establish a correspondence between an external magnetic flux and internal fictitious fluxes of topological defects in a Chern insulator on the honeycomb lattice. While the precise correspondence holds for a specific model on the honeycomb lattice, our edge-state picture suggests similar results for other topological models with crystalline symmetry (including topological superconductors), in line with Ref.~\cite{Teo:2012}. Our results also generalize to time-reversal invariant TIs. The disclinations then act as a source of spin-flux \cite{Qi:2008}, i.e.~a flux with opposite sign for the two spin components. The spectrum induced by the topological defects can be measured by scanning tunneling microscopy offering a probe of the topological state which is complementary to measuring the quantized edge currents.

\acknowledgements
We are grateful to F.~de Juan, A.~H.~MacDonald, J.~E.~Moore, Q.~Niu, A.~Vishwanath and P.~Wiegmann for discussions and helpful comments. AR acknowledges support through the Swiss National Science Foundation.
\bibliography{biblio}

\begin{thebibliography}{10}%
\makeatletter
\providecommand \@ifxundefined [1]{%
 \ifx #1\undefined \expandafter \@firstoftwo
 \else \expandafter \@secondoftwo
\fi
}%
\providecommand \@ifnum [1]{%
 \ifnum #1\expandafter \@firstoftwo
 \else \expandafter \@secondoftwo
\fi
}%
\providecommand \enquote [1]{``#1''}%
\providecommand \bibnamefont  [1]{#1}%
\providecommand \bibfnamefont [1]{#1}%
\providecommand \citenamefont [1]{#1}%
\providecommand\href[0]{\@sanitize\@href}%
\providecommand\@href[1]{\endgroup\@@startlink{#1}\endgroup\@@href}%
\providecommand\@@href[1]{#1\@@endlink}%
\providecommand \@sanitize [0]{\begingroup\catcode`\&12\catcode`\#12\relax}%
\@ifxundefined \pdfoutput {\@firstoftwo}{%
 \@ifnum{\z@=\pdfoutput}{\@firstoftwo}{\@secondoftwo}%
}{%
 \providecommand\@@startlink[1]{\leavevmode\special{html:<a href="#1">}}%
 \providecommand\@@endlink[0]{\special{html:</a>}}%
}{%
 \providecommand\@@startlink[1]{%
  \leavevmode
  \pdfstartlink
   attr{/Border[0 0 1 ]/H/I/C[0 1 1]}%
   user{/Subtype/Link/A<</Type/Action/S/URI/URI(#1)>>}%
  \relax
 }%
 \providecommand\@@endlink[0]{\pdfendlink}%
}%
\providecommand \url  [0]{\begingroup\@sanitize \@url }%
\providecommand \@url [1]{\endgroup\@href {#1}{\urlprefix}}%
\providecommand \urlprefix [0]{URL }%
\providecommand \Eprint[0]{\href }%
\@ifxundefined \urlstyle {%
  \providecommand \doi [1]{doi:\discretionary{}{}{}#1}%
}{%
  \providecommand \doi [0]{doi:\discretionary{}{}{}\begingroup
  \urlstyle{rm}\Url }%
}%
\providecommand \doibase [0]{http://dx.doi.org/}%
\providecommand \Doi[1]{\href{\doibase#1}}%
\providecommand \bibAnnote [3]{%
  \BibitemShut{#1}%
  \begin{quotation}\noindent
    \textsc{Key:}\ #2\\\textsc{Annotation:}\ #3%
  \end{quotation}%
}%
\providecommand \bibAnnoteFile [2]{%
  \IfFileExists{#2}{\bibAnnote {#1} {#2} {\input{#2}}}{}%
}%
\providecommand \typeout [0]{\immediate \write \m@ne }%
\providecommand \selectlanguage [0]{\@gobble}%
\providecommand \bibinfo [0]{\@secondoftwo}%
\providecommand \bibfield [0]{\@secondoftwo}%
\providecommand \translation [1]{[#1]}%
\providecommand \BibitemOpen[0]{}%
\providecommand \bibitemStop [0]{}%
\providecommand \bibitemNoStop [0]{.\EOS\space}%
\providecommand \EOS [0]{\spacefactor3000\relax}%
\providecommand \BibitemShut [1]{\csname bibitem#1\endcsname}%
\bibitem{Moore:2010}%
  \BibitemOpen
  \bibfield{author}{%
  \bibinfo {author} {\bibfnamefont{J.~E.}\ \bibnamefont{Moore}},\ }%
  \bibfield{journal}{%
  \bibinfo {journal} {Nature}\ }%
  \textbf{\bibinfo {volume} {464}},\ \bibinfo {pages} {194} (\bibinfo {month}
  {03}\ \bibinfo {year} {2010})%
  \bibAnnoteFile{NoStop}{Moore:2010}%
\bibitem{Hasan:2010}%
  \BibitemOpen
  \bibfield{author}{%
  \bibinfo {author} {\bibfnamefont{M.~Z.}\ \bibnamefont{Hasan}}\ and\ \bibinfo
  {author} {\bibfnamefont{C.~L.}\ \bibnamefont{Kane}},\ }%
  \bibfield{journal}{%
  \Doi{10.1103/RevModPhys.82.3045}{\bibinfo {journal} {Rev. Mod. Phys.}}\ }%
  \textbf{\bibinfo {volume} {82}},\ \bibinfo {pages} {3045} (\bibinfo {year}
  {2010})%
  \bibAnnoteFile{NoStop}{Hasan:2010}%
\bibitem{Qi:2011b}%
  \BibitemOpen
  \bibfield{author}{%
  \bibinfo {author} {\bibfnamefont{X.-L.}\ \bibnamefont{Qi}}\ and\ \bibinfo
  {author} {\bibfnamefont{S.-C.}\ \bibnamefont{Zhang}},\ }%
  \bibfield{journal}{%
  \Doi{10.1103/RevModPhys.83.1057}{\bibinfo {journal} {Rev. Mod. Phys.}}\ }%
  \textbf{\bibinfo {volume} {83}},\ \bibinfo {pages} {1057} (\bibinfo {year}
  {2011})%
  \bibAnnoteFile{NoStop}{Qi:2011b}%
\bibitem{Mong:2010}%
  \BibitemOpen
  \bibfield{author}{%
  \bibinfo {author} {\bibfnamefont{R.~S.~K.}\ \bibnamefont{Mong}}, \bibinfo
  {author} {\bibfnamefont{A.~M.}\ \bibnamefont{Essin}},\ and\ \bibinfo {author}
  {\bibfnamefont{J.~E.}\ \bibnamefont{Moore}},\ }%
  \bibfield{journal}{%
  \Doi{10.1103/PhysRevB.81.245209}{\bibinfo {journal} {Phys. Rev. B}}\ }%
  \textbf{\bibinfo {volume} {81}},\ \bibinfo {pages} {245209} (\bibinfo {year}
  {2010})%
  \bibAnnoteFile{NoStop}{Mong:2010}%
\bibitem{Hughes:2011}%
  \BibitemOpen
  \bibfield{author}{%
  \bibinfo {author} {\bibfnamefont{T.~L.}\ \bibnamefont{Hughes}}, \bibinfo
  {author} {\bibfnamefont{E.}~\bibnamefont{Prodan}},\ and\ \bibinfo {author}
  {\bibfnamefont{B.~A.}\ \bibnamefont{Bernevig}},\ }%
  \bibfield{journal}{%
  \Doi{10.1103/PhysRevB.83.245132}{\bibinfo {journal} {Phys. Rev. B}}\ }%
  \textbf{\bibinfo {volume} {83}},\ \bibinfo {pages} {245132} (\bibinfo {year}
  {2011})%
  \bibAnnoteFile{NoStop}{Hughes:2011}%
\bibitem{Fu:2011}%
  \BibitemOpen
  \bibfield{author}{%
  \bibinfo {author} {\bibfnamefont{L.}~\bibnamefont{Fu}},\ }%
  \bibfield{journal}{%
  \Doi{10.1103/PhysRevLett.106.106802}{\bibinfo {journal} {Phys. Rev. Lett.}}\
  }%
  \textbf{\bibinfo {volume} {106}},\ \bibinfo {pages} {106802} (\bibinfo {year}
  {2011})%
  \bibAnnoteFile{NoStop}{Fu:2011}%
\bibitem{Dziawa:2012}%
  \BibitemOpen
  \bibfield{author}{%
  \bibinfo {author} {\bibfnamefont{P.}~\bibnamefont{Dziawa}}, \bibinfo {author}
  {\bibfnamefont{B.~J.}\ \bibnamefont{Kowalski}}, \bibinfo {author}
  {\bibfnamefont{K.}~\bibnamefont{Dybko}}, \bibinfo {author}
  {\bibfnamefont{R.}~\bibnamefont{Buczko}}, \bibinfo {author}
  {\bibfnamefont{A.}~\bibnamefont{Szczerbakow}}, \bibinfo {author}
  {\bibfnamefont{M.}~\bibnamefont{Szot}}, \bibinfo {author}
  {\bibfnamefont{E.}~\bibnamefont{{\AA}usakowska}}, \bibinfo {author}
  {\bibfnamefont{T.}~\bibnamefont{Balasubramanian}}, \bibinfo {author}
  {\bibfnamefont{B.~M.}\ \bibnamefont{Wojek}}, \bibinfo {author}
  {\bibfnamefont{M.~H.}\ \bibnamefont{Berntsen}}, \bibinfo {author}
  {\bibfnamefont{O.}~\bibnamefont{Tjernberg}},\ and\ \bibinfo {author}
  {\bibfnamefont{T.}~\bibnamefont{Story}},\ }%
  \bibfield{journal}{%
  \Doi{10.1038/nmat3449}{\bibinfo {journal} {Nat Mater}}\ }%
  \textbf{\bibinfo {volume} {advance online publication}},\  (\bibinfo {month}
  {09}\ \bibinfo {year} {2012})%
  \bibAnnoteFile{NoStop}{Dziawa:2012}%
\bibitem{Xu:2012}%
  \BibitemOpen
  \bibfield{author}{%
  \bibinfo {author} {\bibfnamefont{S.-Y.}\ \bibnamefont{{Xu}}}, \bibinfo
  {author} {\bibfnamefont{C.}~\bibnamefont{{Liu}}}, \bibinfo {author}
  {\bibfnamefont{N.}~\bibnamefont{{Alidoust}}}, \bibinfo {author}
  {\bibfnamefont{D.}~\bibnamefont{{Qian}}}, \bibinfo {author}
  {\bibfnamefont{M.}~\bibnamefont{{Neupane}}}, \bibinfo {author}
  {\bibfnamefont{J.~D.}\ \bibnamefont{{Denlinger}}}, \bibinfo {author}
  {\bibfnamefont{Y.~J.}\ \bibnamefont{{Wang}}}, \bibinfo {author}
  {\bibfnamefont{L.~A.}\ \bibnamefont{{Wray}}}, \bibinfo {author}
  {\bibfnamefont{R.~J.}\ \bibnamefont{{Cava}}}, \bibinfo {author}
  {\bibfnamefont{H.}~\bibnamefont{{Lin}}}, \bibinfo {author}
  {\bibfnamefont{A.}~\bibnamefont{{Marcinkova}}}, \bibinfo {author}
  {\bibfnamefont{E.}~\bibnamefont{{Morosan}}}, \bibinfo {author}
  {\bibfnamefont{A.}~\bibnamefont{{Bansil}}},\ and\ \bibinfo {author}
  {\bibfnamefont{M.~Z.}\ \bibnamefont{{Hasan}}},\ }%
  \bibfield{journal}{%
  \bibinfo {journal} {ArXiv e-prints}}%
   (\bibinfo {month} {Jun.}\ \bibinfo {year} {2012}),\
  \Eprint{http://arxiv.org/abs/1206.2088}{arXiv:1206.2088 [cond-mat.mes-hall]}%
  \bibAnnoteFile{NoStop}{Xu:2012}%
\bibitem{Ran:2009}%
  \BibitemOpen
  \bibfield{author}{%
  \bibinfo {author} {\bibfnamefont{Y.}~\bibnamefont{Ran}}, \bibinfo {author}
  {\bibfnamefont{Y.}~\bibnamefont{Zhang}},\ and\ \bibinfo {author}
  {\bibfnamefont{A.}~\bibnamefont{Vishwanath}},\ }%
  \bibfield{journal}{%
  \Doi{10.1038/nphys1220}{\bibinfo {journal} {Nat Phys}}\ }%
  \textbf{\bibinfo {volume} {5}},\ \bibinfo {pages} {298} (\bibinfo {year}
  {2009})%
  \bibAnnoteFile{NoStop}{Ran:2009}%
\bibitem{Teo:2010}%
  \BibitemOpen
  \bibfield{author}{%
  \bibinfo {author} {\bibfnamefont{J.~C.~Y.}\ \bibnamefont{Teo}}\ and\ \bibinfo
  {author} {\bibfnamefont{C.~L.}\ \bibnamefont{Kane}},\ }%
  \bibfield{journal}{%
  \Doi{10.1103/PhysRevB.82.115120}{\bibinfo {journal} {Phys. Rev. B}}\ }%
  \textbf{\bibinfo {volume} {82}},\ \bibinfo {pages} {115120} (\bibinfo {year}
  {2010})%
  \bibAnnoteFile{NoStop}{Teo:2010}%
\bibitem{Ran:2010}%
  \BibitemOpen
  \bibfield{author}{%
  \bibinfo {author} {\bibfnamefont{Y.}~\bibnamefont{{Ran}}},\ }%
  \bibfield{journal}{%
  \bibinfo {journal} {ArXiv e-prints}}%
   (\bibinfo {year} {2010}),\
  \Eprint{http://arxiv.org/abs/1006.5454}{arXiv:1006.5454}%
  \bibAnnoteFile{NoStop}{Ran:2010}%
\bibitem{Juricic:2012}%
  \BibitemOpen
  \bibfield{author}{%
  \bibinfo {author} {\bibfnamefont{V.}~\bibnamefont{Juri\ifmmode \check{c}\else
  \v{c}\fi{}i\ifmmode~\acute{c}\else \'{c}\fi{}}}, \bibinfo {author}
  {\bibfnamefont{A.}~\bibnamefont{Mesaros}}, \bibinfo {author}
  {\bibfnamefont{R.-J.}\ \bibnamefont{Slager}},\ and\ \bibinfo {author}
  {\bibfnamefont{J.}~\bibnamefont{Zaanen}},\ }%
  \bibfield{journal}{%
  \Doi{10.1103/PhysRevLett.108.106403}{\bibinfo {journal} {Phys. Rev. Lett.}}\
  }%
  \textbf{\bibinfo {volume} {108}},\ \bibinfo {pages} {106403} (\bibinfo {year}
  {2012})%
  \bibAnnoteFile{NoStop}{Juricic:2012}%
\bibitem{Lee:2007}%
  \BibitemOpen
  \bibfield{author}{%
  \bibinfo {author} {\bibfnamefont{D.-H.}\ \bibnamefont{Lee}}, \bibinfo
  {author} {\bibfnamefont{G.-M.}\ \bibnamefont{Zhang}},\ and\ \bibinfo {author}
  {\bibfnamefont{T.}~\bibnamefont{Xiang}},\ }%
  \bibfield{journal}{%
  \Doi{10.1103/PhysRevLett.99.196805}{\bibinfo {journal} {Phys. Rev. Lett.}}\
  }%
  \textbf{\bibinfo {volume} {99}},\ \bibinfo {pages} {196805} (\bibinfo {year}
  {2007})%
  \bibAnnoteFile{NoStop}{Lee:2007}%
\bibitem{Ran:2008}%
  \BibitemOpen
  \bibfield{author}{%
  \bibinfo {author} {\bibfnamefont{Y.}~\bibnamefont{Ran}}, \bibinfo {author}
  {\bibfnamefont{A.}~\bibnamefont{Vishwanath}},\ and\ \bibinfo {author}
  {\bibfnamefont{D.-H.}\ \bibnamefont{Lee}},\ }%
  \bibfield{journal}{%
  \Doi{10.1103/PhysRevLett.101.086801}{\bibinfo {journal} {Phys. Rev. Lett.}}\
  }%
  \textbf{\bibinfo {volume} {101}},\ \bibinfo {pages} {086801} (\bibinfo {year}
  {2008})%
  \bibAnnoteFile{NoStop}{Ran:2008}%
\bibitem{Qi:2008}%
  \BibitemOpen
  \bibfield{author}{%
  \bibinfo {author} {\bibfnamefont{X.-L.}\ \bibnamefont{Qi}}\ and\ \bibinfo
  {author} {\bibfnamefont{S.-C.}\ \bibnamefont{Zhang}},\ }%
  \bibfield{journal}{%
  \Doi{10.1103/PhysRevLett.101.086802}{\bibinfo {journal} {Phys. Rev. Lett.}}\
  }%
  \textbf{\bibinfo {volume} {101}},\ \bibinfo {pages} {086802} (\bibinfo {year}
  {2008})%
  \bibAnnoteFile{NoStop}{Qi:2008}%
\bibitem{Ruegg:2012}%
  \BibitemOpen
  \bibfield{author}{%
  \bibinfo {author} {\bibfnamefont{A.}~\bibnamefont{R\"uegg}}\ and\ \bibinfo
  {author} {\bibfnamefont{G.~A.}\ \bibnamefont{Fiete}},\ }%
  \bibfield{journal}{%
  \Doi{10.1103/PhysRevLett.108.046401}{\bibinfo {journal} {Phys. Rev. Lett.}}\
  }%
  \textbf{\bibinfo {volume} {108}},\ \bibinfo {pages} {046401} (\bibinfo {year}
  {2012})%
  \bibAnnoteFile{NoStop}{Ruegg:2012}%
\bibitem{Assad:2012}%
  \BibitemOpen
  \bibfield{author}{%
  \bibinfo {author} {\bibfnamefont{F.~F.}\ \bibnamefont{{Assaad}}}, \bibinfo
  {author} {\bibfnamefont{M.}~\bibnamefont{{Bercx}}},\ and\ \bibinfo {author}
  {\bibfnamefont{M.}~\bibnamefont{{Hohenadler}}},\ }%
  \bibfield{journal}{%
  \bibinfo {journal} {ArXiv e-prints}}%
   (\bibinfo {month} {Apr.}\ \bibinfo {year} {2012}),\
  \Eprint{http://arxiv.org/abs/1204.4728}{arXiv:1204.4728 [cond-mat.str-el]}%
  \bibAnnoteFile{NoStop}{Assad:2012}%
\bibitem{Zhang:2009}%
  \BibitemOpen
  \bibfield{author}{%
  \bibinfo {author} {\bibfnamefont{Y.}~\bibnamefont{Zhang}}, \bibinfo {author}
  {\bibfnamefont{Y.}~\bibnamefont{Ran}},\ and\ \bibinfo {author}
  {\bibfnamefont{A.}~\bibnamefont{Vishwanath}},\ }%
  \bibfield{journal}{%
  \Doi{10.1103/PhysRevB.79.245331}{\bibinfo {journal} {Phys. Rev. B}}\ }%
  \textbf{\bibinfo {volume} {79}},\ \bibinfo {eid} {245331} (\bibinfo {year}
  {2009})%
  \bibAnnoteFile{NoStop}{Zhang:2009}%
\bibitem{Rosenberg:2010b}%
  \BibitemOpen
  \bibfield{author}{%
  \bibinfo {author} {\bibfnamefont{G.}~\bibnamefont{Rosenberg}}, \bibinfo
  {author} {\bibfnamefont{H.-M.}\ \bibnamefont{Guo}},\ and\ \bibinfo {author}
  {\bibfnamefont{M.}~\bibnamefont{Franz}},\ }%
  \bibfield{journal}{%
  \Doi{10.1103/PhysRevB.82.041104}{\bibinfo {journal} {Phys. Rev. B}}\ }%
  \textbf{\bibinfo {volume} {82}},\ \bibinfo {pages} {041104} (\bibinfo {year}
  {2010})%
  \bibAnnoteFile{NoStop}{Rosenberg:2010b}%
\bibitem{Peng:2010}%
  \BibitemOpen
  \bibfield{author}{%
  \bibinfo {author} {\bibfnamefont{H.}~\bibnamefont{Peng}}, \bibinfo {author}
  {\bibfnamefont{K.}~\bibnamefont{Lai}}, \bibinfo {author}
  {\bibfnamefont{D.}~\bibnamefont{Kong}}, \bibinfo {author}
  {\bibfnamefont{S.}~\bibnamefont{Meister}}, \bibinfo {author}
  {\bibfnamefont{Y.}~\bibnamefont{Chen}}, \bibinfo {author}
  {\bibfnamefont{X.-L.}\ \bibnamefont{Qi}}, \bibinfo {author}
  {\bibfnamefont{S.-C.}\ \bibnamefont{Zhang}}, \bibinfo {author}
  {\bibfnamefont{Z.-X.}\ \bibnamefont{Shen}},\ and\ \bibinfo {author}
  {\bibfnamefont{Y.}~\bibnamefont{Cui}},\ }%
  \bibfield{journal}{%
  \Doi{10.1038/nmat2609}{\bibinfo {journal} {Nat Mater}}\ }%
  \textbf{\bibinfo {volume} {9}},\ \bibinfo {pages} {225} (\bibinfo {month}
  {03}\ \bibinfo {year} {2010})%
  \bibAnnoteFile{NoStop}{Peng:2010}%
\bibitem{Bardarson:2010}%
  \BibitemOpen
  \bibfield{author}{%
  \bibinfo {author} {\bibfnamefont{J.~H.}\ \bibnamefont{Bardarson}}, \bibinfo
  {author} {\bibfnamefont{P.~W.}\ \bibnamefont{Brouwer}},\ and\ \bibinfo
  {author} {\bibfnamefont{J.~E.}\ \bibnamefont{Moore}},\ }%
  \bibfield{journal}{%
  \Doi{10.1103/PhysRevLett.105.156803}{\bibinfo {journal} {Phys. Rev. Lett.}}\
  }%
  \textbf{\bibinfo {volume} {105}},\ \bibinfo {pages} {156803} (\bibinfo {year}
  {2010})%
  \bibAnnoteFile{NoStop}{Bardarson:2010}%
\bibitem{ZhangYi:2010}%
  \BibitemOpen
  \bibfield{author}{%
  \bibinfo {author} {\bibfnamefont{Y.}~\bibnamefont{Zhang}}\ and\ \bibinfo
  {author} {\bibfnamefont{A.}~\bibnamefont{Vishwanath}},\ }%
  \bibfield{journal}{%
  \Doi{10.1103/PhysRevLett.105.206601}{\bibinfo {journal} {Phys. Rev. Lett.}}\
  }%
  \textbf{\bibinfo {volume} {105}},\ \bibinfo {pages} {206601} (\bibinfo {year}
  {2010})%
  \bibAnnoteFile{NoStop}{ZhangYi:2010}%
\bibitem{Gonzalez:1993}%
  \BibitemOpen
  \bibfield{author}{%
  \bibinfo {author} {\bibfnamefont{J.}~\bibnamefont{Gonz{\'a}lez}}, \bibinfo
  {author} {\bibfnamefont{F.}~\bibnamefont{Guinea}},\ and\ \bibinfo {author}
  {\bibfnamefont{M.~A.~H.}\ \bibnamefont{Vozmediano}},\ }%
  \bibfield{journal}{%
  \Doi{10.1016/0550-3213(93)90009-E}{\bibinfo {journal} {Nuclear Physics B}}\
  }%
  \textbf{\bibinfo {volume} {406}},\ \bibinfo {pages} {771} (\bibinfo {month}
  {10}\ \bibinfo {year} {1993})%
  \bibAnnoteFile{NoStop}{Gonzalez:1993}%
\bibitem{Lammert:2000}%
  \BibitemOpen
  \bibfield{author}{%
  \bibinfo {author} {\bibfnamefont{P.~E.}\ \bibnamefont{Lammert}}\ and\
  \bibinfo {author} {\bibfnamefont{V.~H.}\ \bibnamefont{Crespi}},\ }%
  \bibfield{journal}{%
  \Doi{10.1103/PhysRevLett.85.5190}{\bibinfo {journal} {Phys. Rev. Lett.}}\ }%
  \textbf{\bibinfo {volume} {85}},\ \bibinfo {pages} {5190} (\bibinfo {year}
  {2000})%
  \bibAnnoteFile{NoStop}{Lammert:2000}%
\bibitem{Lammert:2004}%
  \BibitemOpen
  \bibfield{author}{%
  \bibinfo {author} {\bibfnamefont{P.~E.}\ \bibnamefont{Lammert}}\ and\
  \bibinfo {author} {\bibfnamefont{V.~H.}\ \bibnamefont{Crespi}},\ }%
  \bibfield{journal}{%
  \Doi{10.1103/PhysRevB.69.035406}{\bibinfo {journal} {Phys. Rev. B}}\ }%
  \textbf{\bibinfo {volume} {69}},\ \bibinfo {pages} {035406} (\bibinfo {year}
  {2004})%
  \bibAnnoteFile{NoStop}{Lammert:2004}%
\bibitem{Cortijo:2007}%
  \BibitemOpen
  \bibfield{author}{%
  \bibinfo {author} {\bibfnamefont{A.}~\bibnamefont{Cortijo}}\ and\ \bibinfo
  {author} {\bibfnamefont{M.}~\bibnamefont{a~A.H.~Vozmediano}},\ }%
  \bibfield{journal}{%
  \Doi{10.1016/j.nuclphysb.2006.10.031}{\bibinfo {journal} {Nuclear Physics
  B}}\ }%
  \textbf{\bibinfo {volume} {763}},\ \bibinfo {pages} {293 } (\bibinfo {year}
  {2007})%
  \bibAnnoteFile{NoStop}{Cortijo:2007}%
\bibitem{Yazyev:2010}%
  \BibitemOpen
  \bibfield{author}{%
  \bibinfo {author} {\bibfnamefont{O.~V.}\ \bibnamefont{Yazyev}}\ and\ \bibinfo
  {author} {\bibfnamefont{S.~G.}\ \bibnamefont{Louie}},\ }%
  \bibfield{journal}{%
  \Doi{10.1103/PhysRevB.81.195420}{\bibinfo {journal} {Phys. Rev. B}}\ }%
  \textbf{\bibinfo {volume} {81}},\ \bibinfo {pages} {195420} (\bibinfo {year}
  {2010})%
  \bibAnnoteFile{NoStop}{Yazyev:2010}%
\bibitem{Cockayne:2011}%
  \BibitemOpen
  \bibfield{author}{%
  \bibinfo {author} {\bibfnamefont{E.}~\bibnamefont{Cockayne}}, \bibinfo
  {author} {\bibfnamefont{G.~M.}\ \bibnamefont{Rutter}}, \bibinfo {author}
  {\bibfnamefont{N.~P.}\ \bibnamefont{Guisinger}}, \bibinfo {author}
  {\bibfnamefont{J.~N.}\ \bibnamefont{Crain}}, \bibinfo {author}
  {\bibfnamefont{P.~N.}\ \bibnamefont{First}},\ and\ \bibinfo {author}
  {\bibfnamefont{J.~A.}\ \bibnamefont{Stroscio}},\ }%
  \bibfield{journal}{%
  \Doi{10.1103/PhysRevB.83.195425}{\bibinfo {journal} {Phys. Rev. B}}\ }%
  \textbf{\bibinfo {volume} {83}},\ \bibinfo {pages} {195425} (\bibinfo {year}
  {2011})%
  \bibAnnoteFile{NoStop}{Cockayne:2011}%
\bibitem{Kane:2005a}%
  \BibitemOpen
  \bibfield{author}{%
  \bibinfo {author} {\bibfnamefont{C.~L.}\ \bibnamefont{Kane}}\ and\ \bibinfo
  {author} {\bibfnamefont{E.~J.}\ \bibnamefont{Mele}},\ }%
  \bibfield{journal}{%
  \Doi{10.1103/PhysRevLett.95.226801}{\bibinfo {journal} {Phys. Rev. Lett.}}\
  }%
  \textbf{\bibinfo {volume} {95}},\ \bibinfo {pages} {226801} (\bibinfo {year}
  {2005})%
  \bibAnnoteFile{NoStop}{Kane:2005a}%
\bibitem{Kane:2005b}%
  \BibitemOpen
  \bibfield{author}{%
  \bibinfo {author} {\bibfnamefont{C.~L.}\ \bibnamefont{Kane}}\ and\ \bibinfo
  {author} {\bibfnamefont{E.~J.}\ \bibnamefont{Mele}},\ }%
  \bibfield{journal}{%
  \Doi{10.1103/PhysRevLett.95.146802}{\bibinfo {journal} {Phys. Rev. Lett.}}\
  }%
  \textbf{\bibinfo {volume} {95}},\ \bibinfo {pages} {146802} (\bibinfo {year}
  {2005})%
  \bibAnnoteFile{NoStop}{Kane:2005b}%
\bibitem{CastroNeto:2009}%
  \BibitemOpen
  \bibfield{author}{%
  \bibinfo {author} {\bibfnamefont{A.~H.}\ \bibnamefont{Castro~Neto}}\ and\
  \bibinfo {author} {\bibfnamefont{F.}~\bibnamefont{Guinea}},\ }%
  \bibfield{journal}{%
  \Doi{10.1103/PhysRevLett.103.026804}{\bibinfo {journal} {Phys. Rev. Lett.}}\
  }%
  \textbf{\bibinfo {volume} {103}},\ \bibinfo {pages} {026804} (\bibinfo {year}
  {2009})%
  \bibAnnoteFile{NoStop}{CastroNeto:2009}%
\bibitem{Weeks:2011}%
  \BibitemOpen
  \bibfield{author}{%
  \bibinfo {author} {\bibfnamefont{C.}~\bibnamefont{Weeks}}, \bibinfo {author}
  {\bibfnamefont{J.}~\bibnamefont{Hu}}, \bibinfo {author}
  {\bibfnamefont{J.}~\bibnamefont{Alicea}}, \bibinfo {author}
  {\bibfnamefont{M.}~\bibnamefont{Franz}},\ and\ \bibinfo {author}
  {\bibfnamefont{R.}~\bibnamefont{Wu}},\ }%
  \bibfield{journal}{%
  \Doi{10.1103/PhysRevX.1.021001}{\bibinfo {journal} {Phys. Rev. X}}\ }%
  \textbf{\bibinfo {volume} {1}},\ \bibinfo {pages} {021001} (\bibinfo {year}
  {2011})%
  \bibAnnoteFile{NoStop}{Weeks:2011}%
\bibitem{Ding:2011}%
  \BibitemOpen
  \bibfield{author}{%
  \bibinfo {author} {\bibfnamefont{J.}~\bibnamefont{Ding}}, \bibinfo {author}
  {\bibfnamefont{Z.}~\bibnamefont{Qiao}}, \bibinfo {author}
  {\bibfnamefont{W.}~\bibnamefont{Feng}}, \bibinfo {author}
  {\bibfnamefont{Y.}~\bibnamefont{Yao}},\ and\ \bibinfo {author}
  {\bibfnamefont{Q.}~\bibnamefont{Niu}},\ }%
  \bibfield{journal}{%
  \Doi{10.1103/PhysRevB.84.195444}{\bibinfo {journal} {Phys. Rev. B}}\ }%
  \textbf{\bibinfo {volume} {84}},\ \bibinfo {pages} {195444} (\bibinfo {year}
  {2011})%
  \bibAnnoteFile{NoStop}{Ding:2011}%
\bibitem{Hu:2012}%
  \BibitemOpen
  \bibfield{author}{%
  \bibinfo {author} {\bibfnamefont{J.}~\bibnamefont{{Hu}}}, \bibinfo {author}
  {\bibfnamefont{J.}~\bibnamefont{{Alicea}}}, \bibinfo {author}
  {\bibfnamefont{R.}~\bibnamefont{{Wu}}},\ and\ \bibinfo {author}
  {\bibfnamefont{M.}~\bibnamefont{{Franz}}},\ }%
  \bibfield{journal}{%
  \bibinfo {journal} {ArXiv e-prints}}%
   (\bibinfo {year} {2012}),\
  \Eprint{http://arxiv.org/abs/1206.4320}{arXiv:1206.4320 [cond-mat.mes-hall]}%
  \bibAnnoteFile{NoStop}{Hu:2012}%
\bibitem{Tse:2011}%
  \BibitemOpen
  \bibfield{author}{%
  \bibinfo {author} {\bibfnamefont{W.-K.}\ \bibnamefont{Tse}}, \bibinfo
  {author} {\bibfnamefont{Z.}~\bibnamefont{Qiao}}, \bibinfo {author}
  {\bibfnamefont{Y.}~\bibnamefont{Yao}}, \bibinfo {author}
  {\bibfnamefont{A.~H.}\ \bibnamefont{MacDonald}},\ and\ \bibinfo {author}
  {\bibfnamefont{Q.}~\bibnamefont{Niu}},\ }%
  \bibfield{journal}{%
  \Doi{10.1103/PhysRevB.83.155447}{\bibinfo {journal} {Phys. Rev. B}}\ }%
  \textbf{\bibinfo {volume} {83}},\ \bibinfo {pages} {155447} (\bibinfo {year}
  {2011})%
  \bibAnnoteFile{NoStop}{Tse:2011}%
\bibitem{Zhenhua:2011}%
  \BibitemOpen
  \bibfield{author}{%
  \bibinfo {author} {\bibfnamefont{Z.}~\bibnamefont{Qiao}}, \bibinfo {author}
  {\bibfnamefont{W.-K.}\ \bibnamefont{Tse}}, \bibinfo {author}
  {\bibfnamefont{H.}~\bibnamefont{Jiang}}, \bibinfo {author}
  {\bibfnamefont{Y.}~\bibnamefont{Yao}},\ and\ \bibinfo {author}
  {\bibfnamefont{Q.}~\bibnamefont{Niu}},\ }%
  \bibfield{journal}{%
  \Doi{10.1103/PhysRevLett.107.256801}{\bibinfo {journal} {Phys. Rev. Lett.}}\
  }%
  \textbf{\bibinfo {volume} {107}},\ \bibinfo {pages} {256801} (\bibinfo {year}
  {2011})%
  \bibAnnoteFile{NoStop}{Zhenhua:2011}%
\bibitem{Gomes:2012}%
  \BibitemOpen
  \bibfield{author}{%
  \bibinfo {author} {\bibfnamefont{K.~K.}\ \bibnamefont{Gomes}}, \bibinfo
  {author} {\bibfnamefont{W.}~\bibnamefont{Mar}}, \bibinfo {author}
  {\bibfnamefont{W.}~\bibnamefont{Ko}}, \bibinfo {author}
  {\bibfnamefont{F.}~\bibnamefont{Guinea}},\ and\ \bibinfo {author}
  {\bibfnamefont{H.~C.}\ \bibnamefont{Manoharan}},\ }%
  \bibfield{journal}{%
  \Doi{10.1038/nature10941}{\bibinfo {journal} {Nature}}\ }%
  \textbf{\bibinfo {volume} {483}},\ \bibinfo {pages} {306} (\bibinfo {month}
  {03}\ \bibinfo {year} {2012})%
  \bibAnnoteFile{NoStop}{Gomes:2012}%
\bibitem{LiuC:2011}%
  \BibitemOpen
  \bibfield{author}{%
  \bibinfo {author} {\bibfnamefont{C.-C.}\ \bibnamefont{Liu}}, \bibinfo
  {author} {\bibfnamefont{W.}~\bibnamefont{Feng}},\ and\ \bibinfo {author}
  {\bibfnamefont{Y.}~\bibnamefont{Yao}},\ }%
  \bibfield{journal}{%
  \Doi{10.1103/PhysRevLett.107.076802}{\bibinfo {journal} {Phys. Rev. Lett.}}\
  }%
  \textbf{\bibinfo {volume} {107}},\ \bibinfo {pages} {076802} (\bibinfo {year}
  {2011})%
  \bibAnnoteFile{NoStop}{LiuC:2011}%
\bibitem{Vogt:2012}%
  \BibitemOpen
  \bibfield{author}{%
  \bibinfo {author} {\bibfnamefont{P.}~\bibnamefont{Vogt}}, \bibinfo {author}
  {\bibfnamefont{P.}~\bibnamefont{De~Padova}}, \bibinfo {author}
  {\bibfnamefont{C.}~\bibnamefont{Quaresima}}, \bibinfo {author}
  {\bibfnamefont{J.}~\bibnamefont{Avila}}, \bibinfo {author}
  {\bibfnamefont{E.}~\bibnamefont{Frantzeskakis}}, \bibinfo {author}
  {\bibfnamefont{M.~C.}\ \bibnamefont{Asensio}}, \bibinfo {author}
  {\bibfnamefont{A.}~\bibnamefont{Resta}}, \bibinfo {author}
  {\bibfnamefont{B.}~\bibnamefont{Ealet}},\ and\ \bibinfo {author}
  {\bibfnamefont{G.}~\bibnamefont{Le~Lay}},\ }%
  \bibfield{journal}{%
  \Doi{10.1103/PhysRevLett.108.155501}{\bibinfo {journal} {Phys. Rev. Lett.}}\
  }%
  \textbf{\bibinfo {volume} {108}},\ \bibinfo {pages} {155501} (\bibinfo {year}
  {2012})%
  \bibAnnoteFile{NoStop}{Vogt:2012}%
\bibitem{Chaikin:2000}%
  \BibitemOpen
  \bibfield{author}{%
  \bibinfo {author} {\bibfnamefont{P.~M.}\ \bibnamefont{Chaikin}}\ and\
  \bibinfo {author} {\bibfnamefont{T.~C.}\ \bibnamefont{Lubensky}},\ }%
  \emph{\bibinfo {title} {Principles of Condensed Matter Physics}}\ (\bibinfo
  {publisher} {Cambridge University Press},\ \bibinfo {year} {2000})%
  \bibAnnoteFile{NoStop}{Chaikin:2000}%
\bibitem{Haldane:1988}%
  \BibitemOpen
  \bibfield{author}{%
  \bibinfo {author} {\bibfnamefont{F.~D.~M.}\ \bibnamefont{Haldane}},\ }%
  \bibfield{journal}{%
  \Doi{10.1103/PhysRevLett.61.2015}{\bibinfo {journal} {Phys. Rev. Lett.}}\ }%
  \textbf{\bibinfo {volume} {61}},\ \bibinfo {pages} {2015} (\bibinfo {year}
  {1988})%
  \bibAnnoteFile{NoStop}{Haldane:1988}%
\bibitem{Grosso:1985}%
  \BibitemOpen
  \bibfield{author}{%
  \bibinfo {author} {\bibfnamefont{G.}~\bibnamefont{Grosso}}\ and\ \bibinfo
  {author} {\bibfnamefont{G.~P.}\ \bibnamefont{Parravicini}},\ }%
  \bibfield{journal}{%
  \Doi{10.1002/9780470142868.ch3}{\bibinfo {journal} {Adv. Chem. Phys.}}\ }%
  \textbf{\bibinfo {volume} {63}},\ \bibinfo {pages} {81} (\bibinfo {year}
  {1985})%
  \bibAnnoteFile{NoStop}{Grosso:1985}%
\bibitem{note:supplementary}%
  \BibitemOpen
  \bibinfo {note} {See supplementary materials for details of the
  calculation.}%
  \bibAnnoteFile{Stop}{note:supplementary}%
\bibitem{Vozmediano:2010}%
  \BibitemOpen
  \bibfield{author}{%
  \bibinfo {author} {\bibfnamefont{M.}~\bibnamefont{Vozmediano}}, \bibinfo
  {author} {\bibfnamefont{M.}~\bibnamefont{Katsnelson}},\ and\ \bibinfo
  {author} {\bibfnamefont{F.}~\bibnamefont{Guinea}},\ }%
  \bibfield{journal}{%
  \Doi{10.1016/j.physrep.2010.07.003}{\bibinfo {journal} {Physics Reports}}\ }%
  \textbf{\bibinfo {volume} {496}},\ \bibinfo {pages} {109 } (\bibinfo {year}
  {2010})%
  \bibAnnoteFile{NoStop}{Vozmediano:2010}%
\bibitem{Guinea:2010}%
  \BibitemOpen
  \bibfield{author}{%
  \bibinfo {author} {\bibfnamefont{F.}~\bibnamefont{Guinea}}, \bibinfo {author}
  {\bibfnamefont{M.~I.}\ \bibnamefont{Katsnelson}},\ and\ \bibinfo {author}
  {\bibfnamefont{A.~K.}\ \bibnamefont{Geim}},\ }%
  \bibfield{journal}{%
  \Doi{10.1038/nphys1420}{\bibinfo {journal} {Nat Phys}}\ }%
  \textbf{\bibinfo {volume} {6}},\ \bibinfo {pages} {30} (\bibinfo {month}
  {01}\ \bibinfo {year} {2010})%
  \bibAnnoteFile{NoStop}{Guinea:2010}%
\bibitem{Ghaemi:2012}%
  \BibitemOpen
  \bibfield{author}{%
  \bibinfo {author} {\bibfnamefont{P.}~\bibnamefont{Ghaemi}}, \bibinfo {author}
  {\bibfnamefont{J.}~\bibnamefont{Cayssol}}, \bibinfo {author}
  {\bibfnamefont{D.~N.}\ \bibnamefont{Sheng}},\ and\ \bibinfo {author}
  {\bibfnamefont{A.}~\bibnamefont{Vishwanath}},\ }%
  \bibfield{journal}{%
  \Doi{10.1103/PhysRevLett.108.266801}{\bibinfo {journal} {Phys. Rev. Lett.}}\
  }%
  \textbf{\bibinfo {volume} {108}},\ \bibinfo {pages} {266801} (\bibinfo {year}
  {2012})%
  \bibAnnoteFile{NoStop}{Ghaemi:2012}%
\bibitem{Mesaros:2010}%
  \BibitemOpen
  \bibfield{author}{%
  \bibinfo {author} {\bibfnamefont{A.}~\bibnamefont{Mesaros}}, \bibinfo
  {author} {\bibfnamefont{D.}~\bibnamefont{Sadri}},\ and\ \bibinfo {author}
  {\bibfnamefont{J.}~\bibnamefont{Zaanen}},\ }%
  \bibfield{journal}{%
  \Doi{10.1103/PhysRevB.82.073405}{\bibinfo {journal} {Phys. Rev. B}}\ }%
  \textbf{\bibinfo {volume} {82}},\ \bibinfo {pages} {073405} (\bibinfo {year}
  {2010})%
  \bibAnnoteFile{NoStop}{Mesaros:2010}%
\bibitem{note:co-rotating}%
  \BibitemOpen
  \bibinfo {note} {The co-rotating spinor is referred to as the spinor
  expressed in the local frame $({\bs e}_r,{\bs e}_{\phi})$ with unit vectors
  along the radial and azimuthal directions.}%
  \bibAnnoteFile{Stop}{note:co-rotating}%
\bibitem{note:valley_comments}%
  \BibitemOpen
  \bibinfo {note} {As a counter-example, a staggered sublattice mass
  $m\sigma_z$ in the presence of an odd-membered disclination does not allow to
  define two decoupled valleys.}%
  \bibAnnoteFile{Stop}{note:valley_comments}%
\bibitem{Slobodeniuk:2010}%
  \BibitemOpen
  \bibfield{author}{%
  \bibinfo {author} {\bibfnamefont{A.~O.}\ \bibnamefont{Slobodeniuk}}, \bibinfo
  {author} {\bibfnamefont{S.~G.}\ \bibnamefont{Sharapov}},\ and\ \bibinfo
  {author} {\bibfnamefont{V.~M.}\ \bibnamefont{Loktev}},\ }%
  \bibfield{journal}{%
  \Doi{10.1103/PhysRevB.82.075316}{\bibinfo {journal} {Phys. Rev. B}}\ }%
  \textbf{\bibinfo {volume} {82}},\ \bibinfo {pages} {075316} (\bibinfo {year}
  {2010})%
  \bibAnnoteFile{NoStop}{Slobodeniuk:2010}%
\bibitem{Roy:2010}%
  \BibitemOpen
  \bibfield{author}{%
  \bibinfo {author} {\bibfnamefont{A.}~\bibnamefont{Roy}}\ and\ \bibinfo
  {author} {\bibfnamefont{M.}~\bibnamefont{Stone}},\ }%
  \bibfield{journal}{%
  \Doi{10.1088/1751-8113/43/1/015203}{\bibinfo {journal} {Journal of Physics A:
  Mathematical and Theoretical}}\ }%
  \textbf{\bibinfo {volume} {43}},\ \bibinfo {pages} {015203} (\bibinfo {year}
  {2010})%
  \bibAnnoteFile{NoStop}{Roy:2010}%
\bibitem{Berry:1987}%
  \BibitemOpen
  \bibfield{author}{%
  \bibinfo {author} {\bibfnamefont{M.~V.}\ \bibnamefont{Berry}}\ and\ \bibinfo
  {author} {\bibfnamefont{R.~J.}\ \bibnamefont{Mondragon}},\ }%
  \bibfield{journal}{%
  \bibinfo {journal} {Proceedings of the Royal Society of London. A.
  Mathematical and Physical Sciences}\ }%
  \textbf{\bibinfo {volume} {412}},\ \bibinfo {pages} {53} (\bibinfo {year}
  {1987})%
  \bibAnnoteFile{NoStop}{Berry:1987}%
\bibitem{Recher:2007}%
  \BibitemOpen
  \bibfield{author}{%
  \bibinfo {author} {\bibfnamefont{P.}~\bibnamefont{Recher}}, \bibinfo {author}
  {\bibfnamefont{B.}~\bibnamefont{Trauzettel}}, \bibinfo {author}
  {\bibfnamefont{A.}~\bibnamefont{Rycerz}}, \bibinfo {author}
  {\bibfnamefont{Y.~M.}\ \bibnamefont{Blanter}}, \bibinfo {author}
  {\bibfnamefont{C.~W.~J.}\ \bibnamefont{Beenakker}},\ and\ \bibinfo {author}
  {\bibfnamefont{A.~F.}\ \bibnamefont{Morpurgo}},\ }%
  \bibfield{journal}{%
  \Doi{10.1103/PhysRevB.76.235404}{\bibinfo {journal} {Phys. Rev. B}}\ }%
  \textbf{\bibinfo {volume} {76}},\ \bibinfo {pages} {235404} (\bibinfo {year}
  {2007})%
  \bibAnnoteFile{NoStop}{Recher:2007}%
\bibitem{Akhmerov:2008}%
  \BibitemOpen
  \bibfield{author}{%
  \bibinfo {author} {\bibfnamefont{A.~R.}\ \bibnamefont{Akhmerov}}\ and\
  \bibinfo {author} {\bibfnamefont{C.~W.~J.}\ \bibnamefont{Beenakker}},\ }%
  \bibfield{journal}{%
  \Doi{10.1103/PhysRevB.77.085423}{\bibinfo {journal} {Phys. Rev. B}}\ }%
  \textbf{\bibinfo {volume} {77}},\ \bibinfo {pages} {085423} (\bibinfo {year}
  {2008})%
  \bibAnnoteFile{NoStop}{Akhmerov:2008}%
\bibitem{Mudry:2007}%
  \BibitemOpen
  \bibfield{author}{%
  \bibinfo {author} {\bibfnamefont{C.-Y.}\ \bibnamefont{Hou}}, \bibinfo
  {author} {\bibfnamefont{C.}~\bibnamefont{Chamon}},\ and\ \bibinfo {author}
  {\bibfnamefont{C.}~\bibnamefont{Mudry}},\ }%
  \bibfield{journal}{%
  \Doi{10.1103/PhysRevLett.98.186809}{\bibinfo {journal} {Phys. Rev. Lett.}}\
  }%
  \textbf{\bibinfo {volume} {98}},\ \bibinfo {pages} {186809} (\bibinfo {year}
  {2007})%
  \bibAnnoteFile{NoStop}{Mudry:2007}%
\bibitem{note:kekule_mass}%
  \BibitemOpen
  \bibinfo {note} {Apart from the Haldane mass, the inversion-symmetric kekule
  distortion is the only other gap-opening perturbation compatible with the
  defect symmetry for all $n$ \cite{Roy:2010}.}%
  \bibAnnoteFile{Stop}{note:kekule_mass}%
\bibitem{note:internal}%
  \BibitemOpen
  \bibinfo {note} {The argument in the present form assumes no internal degrees
  of freedom, i.e.~each site has only one orbital.}%
  \bibAnnoteFile{Stop}{note:internal}%
\bibitem{Su:1979}%
  \BibitemOpen
  \bibfield{author}{%
  \bibinfo {author} {\bibfnamefont{W.~P.}\ \bibnamefont{Su}}, \bibinfo {author}
  {\bibfnamefont{J.~R.}\ \bibnamefont{Schrieffer}},\ and\ \bibinfo {author}
  {\bibfnamefont{A.~J.}\ \bibnamefont{Heeger}},\ }%
  \bibfield{journal}{%
  \Doi{10.1103/PhysRevLett.42.1698}{\bibinfo {journal} {Phys. Rev. Lett.}}\ }%
  \textbf{\bibinfo {volume} {42}},\ \bibinfo {pages} {1698} (\bibinfo {year}
  {1979})%
  \bibAnnoteFile{NoStop}{Su:1979}%
\bibitem{Teo:2012}%
  \BibitemOpen
  \bibfield{author}{%
  \bibinfo {author} {\bibfnamefont{J.~C.~Y.}\ \bibnamefont{{Teo}}}\ and\
  \bibinfo {author} {\bibfnamefont{T.~L.}\ \bibnamefont{{Hughes}}},\ }%
  \bibfield{journal}{%
  \bibinfo {journal} {ArXiv e-prints}}%
   (\bibinfo {year} {2012}),\
  \Eprint{http://arxiv.org/abs/1208.6303}{arXiv:1208.6303}%
  \bibAnnoteFile{NoStop}{Teo:2012}%
\end{thebibliography}%

\onecolumngrid
\vspace{1in}
\begin{center}
{\bf \large Supplementary materials}
\end{center}
\vspace{0.1in}

\renewcommand{\thetable}{S\Roman{table}}
\renewcommand{\thefigure}{S\arabic{figure}}
\renewcommand{\thesubsection}{S\arabic{subsection}}
\renewcommand{\theequation}{S\arabic{equation}}

\setcounter{secnumdepth}{1}
\setcounter{equation}{0}
\setcounter{figure}{0}
\setcounter{section}{0}

\section{Base functions and boundary conditions}
\begin{figure}[h]
\includegraphics[width=0.5\linewidth]{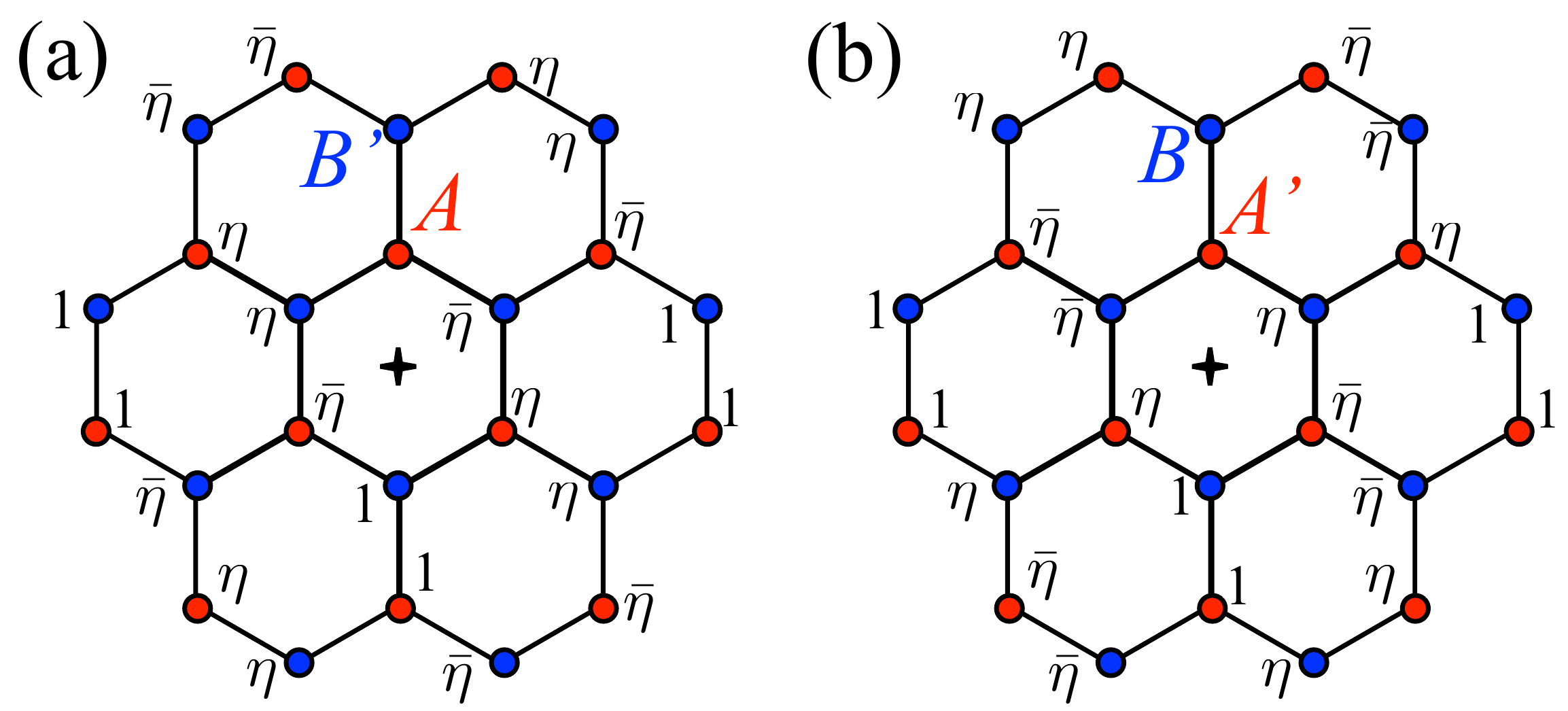}
\caption{The complex phases of the base functions in graphene. $A$ and $B$ refer to the sublattice components at valley ${\bs K}$ while $A'$ and $B'$ refer to the sublattice components at ${\bs K}'$. We use the notation $\eta=e^{2\pi i/3}$ and $\bar{\eta} =e^{-2\pi i/3}$. (a) shows $u_A({\bs r})$ and $u_{B'}({\bs r})$ while (b) shows $u_{A'}({\bs r})$ and $u_{B}({\bs r})$.}
\label{fig:basefunctions}
\end{figure}

The boundary conditions in the cut and glue procedure for the different disclinations are obtained by matching the total wave function across the cut \cite{Lammert:2000}. To derive the continuum theory, this requires to fix a convention for the base function. It is convenient to choose a set of base functions which explicitly preserves the inversion symmetry with respect to the origin located at a center of a hexagon. For valley ${\bs K}=(4\pi/(3\sqrt{3}d),0)$, the base functions are denoted by
$u_A({\bs r})$ and $u_B({\bs r})$; for valley ${\bs K}'=-{\bs K}$ they are denoted by $u_{A'}({\bs r})$ and $u_{B'}({\bs r})$. The complex phases are given by the Bloch factors $e^{\pm i{\bs r}\cdot{\bs K}}$ with ${\bs r}$ the location of the given $A$ or $B$ site. Using the notation 
$\eta=e^{2\pi i/3}$ and $\bar{\eta}=e^{-2\pi i/3}$, the amplitude of the base functions are given in Fig.~\ref{fig:basefunctions}.

The total wave function is then expanded around the four base functions as
\begin{eqnarray}
\psi_{\rm tot}({\bs r})=\psi_A({\bs r})u_A({\bs r})+\psi_B({\bs r})u_B({\bs r})+\psi_{A'}({\bs r})u_{A'}({\bs r})+\psi_{B'}({\bs r})u_{B'}({\bs r}),
\end{eqnarray}
where we have suppressed the dependence on the physical spin. The envelope function 
\begin{equation}
\Psi({\bs r})=[\psi_A({\bs r}),\psi_B({\bs r}),\psi_{A'}({\bs r}),\psi_{B'}({\bs r})]^T
\end{equation} 
is a four-component function which is assumed to be smooth on the atomic scale. In this convention, the Dirac Hamiltonian in the presence of a Haldane mass is given by Eq.~\eqref{eq:Dirac}
\begin{equation}
H=v(\tau_z\sigma_xp_x+\sigma_yp_y)+m\tau_z\sigma_z
\end{equation}
where $\vec{\sigma}=(\sigma_x,\sigma_y,\sigma_z)$ and $\vec{\tau}=(\tau_x,\tau_y,\tau_z)$ are Pauli matrices denoting the sublattice and valley degrees of freedom, respectively.

\begin{figure}
\includegraphics[width=0.6\linewidth]{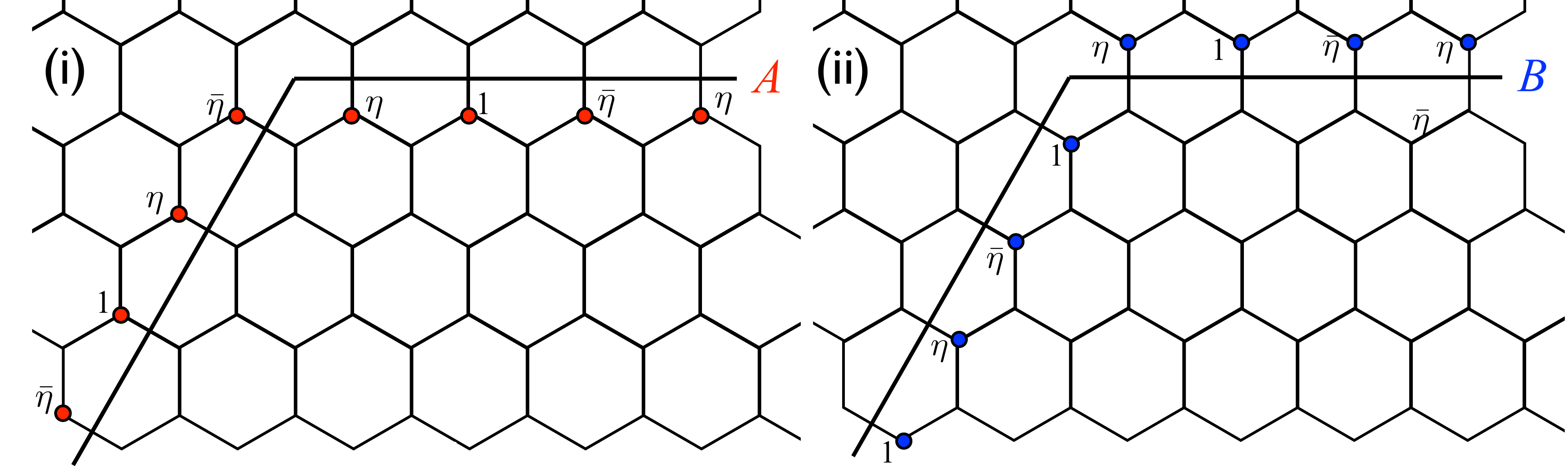}
\caption{The amplitudes of the base functions of valley ${\bs K}$ across the cut for the $120^{\circ}$-disclination. The amplitudes for valley ${\bs K}'$ are obtained by complex conjugation of the above shown amplitudes.}
\label{fig:boundaryconditions120}
\end{figure}

\begin{figure}
\includegraphics[width=0.6\linewidth]{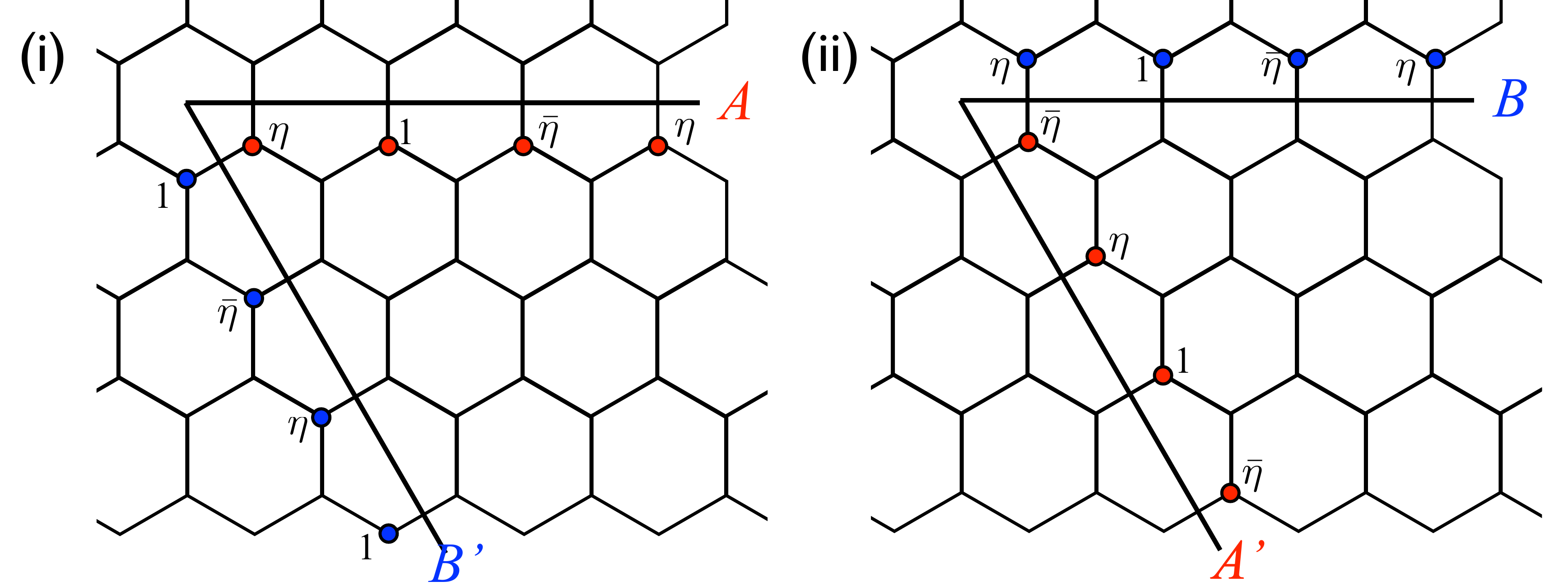}
\caption{The amplitudes of the base functions corresponding to the components which are matched across the cut for the $60^{\circ}$-disclination. The remaining two components are obtained by complex conjugation of the above shown amplitudes.}
\label{fig:boundaryconditions60}
\end{figure}

The boundary conditions of the {\it envelope} function are to make sure that the phase miss-matches of the base functions, as shown in Fig.~\ref{fig:boundaryconditions120} and ~\ref{fig:boundaryconditions60}, are properly compensated so that the {\it total} wave function is single-valued upon encircling the defect core.
For the even-membered disclinations $(n=\pm2)$, the single valuedness of the total wave function implies the following boundary conditions for the envelope function:
\begin{eqnarray}
\Psi(r,\phi=4\pi/3)&=&e^{-i\frac{2\pi}{3}\sigma_z\tau_z}\Psi(r,\phi=0),\quad n=2,\label{eq:BC120}\\
\Psi(r,\phi=8\pi/3)&=&e^{+i\frac{2\pi}{3}\sigma_z\tau_z}\Psi(r,\phi=0),\quad n=-2.
\end{eqnarray}
Note that these boundary conditions are diagonal in sublattice and valley degrees of freedom.

For the $\pm60^{\circ}$-disclinations the sublattice components are no longer conserved. Nevertheless, it is possible to find boundary conditions which are independent of the distance from the defect core by matching opposite sublattice components in opposite valleys, as shown in Fig.~\ref{fig:boundaryconditions60}.
The boundary conditions are then given by
\begin{equation}
\Psi\left(\phi=\frac{5\pi}{3}\right)=
\begin{pmatrix}
0&0&0&\eta\\
0&0&\bar{\eta}&0\\
0&\bar{\eta}&0&0\\
\eta&0&0&0
\end{pmatrix}\Psi(\phi=0),\quad n=1,
\end{equation}
for the pentagon ($+60^{\circ}$) disclination and by
\begin{equation}
\Psi\left(\phi=\frac{7\pi}{3}\right)=
\begin{pmatrix}
0&0&0&\bar{\eta}\\
0&0&\eta&0\\
0&\eta&0&0\\
\bar{\eta}&0&0&0
\end{pmatrix}\Psi(\phi=0),\quad n=-1,
\label{eq:BCm60}
\end{equation}
for the heptagon ($-60^{\circ}$) disclination. 
With the rescaled angular variable $\theta=\phi/\Omega_n$ ($\Omega_n=1-n/6$),
Eq.~\eqref{eq:BC120}-\eqref{eq:BCm60} can be written in a compact form as Eq.~\eqref{eq:UnBC}:
\begin{equation}
\Psi(\theta=2\pi)=e^{i\frac{\pi n}{3}\left(\frac{\sigma_z\tau_z-3\sigma_y\tau_y}{2}\right)}\Psi(\theta=0).
\end{equation}

\section{Block diagonalization of Dirac Hamiltonian}
After performing the transformation to the co-rotating frame [see Eq.~\eqref{eq:Psi_n}], 
\begin{equation}
\begin{pmatrix}
\tilde{H}_K&0\\
0&\tilde{H}_{K'}
\end{pmatrix}=U(\phi)HU^{\dag}(\phi), \quad U(\phi)=e^{i\frac{\phi}{2}\sigma_z\tau_z},
\end{equation}
the Dirac Hamiltonian in polar coordinates reads
\begin{equation}
\tilde{H}_K=
\begin{pmatrix}
m&-i\partial_r-\frac{1}{r}\partial_{\phi}-\frac{i}{r}\left(\frac{\Phi}{\Omega_n\Phi_0}+\frac{1}{2}\right)\\
-i\partial_r+\frac{1}{r}\partial_{\phi}+\frac{i}{r}\left(\frac{\Phi}{\Omega_n\Phi_0}-\frac{1}{2}\right)&-m
\end{pmatrix}.
\label{eq:HK}
\end{equation}
for valley $K$. In the opposite valley, it is given by
\begin{equation}
\tilde{H}_{K'}=
\begin{pmatrix}
-m&i\partial_r-\frac{1}{r}\partial_{\phi}-\frac{i}{r}\left(\frac{\Phi}{\Omega_n\Phi_0}-\frac{1}{2}\right)\\
i\partial_r+\frac{1}{r}\partial_{\phi}+\frac{i}{r}\left(\frac{\Phi}{\Omega_n\Phi_0}+\frac{1}{2}\right)&m
\end{pmatrix}
\label{eq:HKp}
\end{equation}
We have also included a magnetic flux $\Phi$ through the origin, see next section.
We introduce the rescaled angular variable $\theta=\phi/\Omega_n$ ($\Omega_n=1-n/6$), and perform the second local transformation with
\begin{equation}
V_n(\theta)=e^{i\theta\frac{n}{4}\sigma_y\tau_y}=\cos\left(\frac{n\theta}{4}\right)\sigma_0\tau_0+i(\sigma_y\tau_y)\sin\left(\frac{n\theta}{4}\right),
\end{equation}
to obtain 
\begin{equation}
\tilde{H}_n=V_n(\theta)
\begin{pmatrix}
\tilde{H}_K&0\\
0&\tilde{H}_{K'}
\end{pmatrix}
V_n^{\dag}(\theta).
\end{equation}
After these two gauge transformations, the transformed spinor
\begin{equation}
\tilde{\Psi}_n=V_nU\Psi
\end{equation}
now obeys anti-periodic boundary conditions for any $n$,
\begin{equation}
\tilde{\Psi}_n(\theta=2\pi)=-\tilde{\Psi}_n(\theta=0),
\label{eq:ap}
\end{equation}
instead of the awkward condition Eq.~\eqref{eq:UnBC}.
The transformed Hamiltonian explicitly reads
\begin{equation}
\tilde{H}_n=
\begin{pmatrix}
m&-i\partial_r-\frac{1}{r}\partial_{\phi}-\frac{i}{r}\left(\frac{\Phi}{\Omega_n\Phi_0}+\frac{1}{2}\right)&\frac{i}{r\Omega_n}\frac{n}{4}&0\\
-i\partial_r+\frac{1}{r}\partial_{\phi}+\frac{i}{r}\left(\frac{\Phi}{\Omega_n\Phi_0}-\frac{1}{2}\right)&-m&0&\frac{i}{r\Omega_n}\frac{n}{4}\\
\frac{-i}{r\Omega_n}\frac{n}{4}&0&-m&i\partial_r-\frac{1}{r}\partial_{\phi}-\frac{i}{r}\left(\frac{\Phi}{\Omega_n\Phi_0}-\frac{1}{2}\right)\\
0&\frac{-i}{r\Omega_n}\frac{n}{4}&i\partial_r+\frac{1}{r}\partial_{\phi}+\frac{i}{r}\left(\frac{\Phi}{\Omega_n\Phi_0}+\frac{1}{2}\right)&m
\end{pmatrix}.
\end{equation}
It is brought to block-diagonal form by applying the global transformation
\begin{equation}
S\tilde{H}_nS^{\dag},\quad S=\frac{1}{\sqrt{2}}\left(1+i\tau_x\sigma_y\right).
\end{equation}
Separation of angular and radial variables,
\begin{equation}
\Psi'_n(r,\theta)=\chi(r)e^{ij\theta},
\end{equation}
with $j$ half-integer to satisfy Eq.~\eqref{eq:ap}, leads to Eq.~\eqref{eq:Hr}.
\section{Magnetic flux with a disclination}
\subsection{Continuum model}
In the effective Hamiltonian, a magnetic flux $\Phi$ passing through origin is described by
replacing the momentum $\mathbf{p}$ in Eq.~\eqref{eq:Dirac} by the canonical momentum $\mathbf{p}+\mathbf{A}$ with the vector potential 
\begin{equation}
\mathbf{A}(r,\phi) =  \frac{\Phi}{2 \pi} \frac{1}{\Omega_nr} (-\sin \phi, \cos \phi).
\end{equation}
with $\Omega_n=1-n/6$. In polar coordinates, the magnetic flux enters the Hamiltonian as shown in Eqs.~\eqref{eq:HK} and \eqref{eq:HKp}.
\subsection{Lattice model}
In the lattice model, a magnetic flux $\Phi$ through the origin is described by modifying the complex phase of the hoppings according to the Peierls substitution. The choice of complex phases is {\it not} unique (since many vector potentials lead to the same magnetic field), and one simple arrangement is now provided: one draws an arbitrary semi-infinite string starting from the origin, and all hopping intersecting with this string is attached with the phase $\Phi$ or $-\Phi$ ($e=\hbar=1$) via
\begin{equation}
t_{ij} \rightarrow t_{ij} e^{i \Phi \, \text{sign}[ (\mathbf{R}_i- \mathbf{R}_j)\times \mathbf{t} \cdot \hat{z}] } 
\end{equation}
where $t_{ij}$ (generally a complex number) is the hopping amplitude from site $i$ at $\mathbf{R}_i$ to site $j$ at $\mathbf{R}_j$ in the absence of external fluxes, and $\mathbf{t}$ is the tangent of the string at the intersection. In our numerical simulations we choose the semi-infinite string to be a straight line starting from the disclination center and crossing the middle of one of the edges of the core polygon.

\end{document}